\documentclass[pdflatex,sn-nature]{sn-jnl}

\usepackage{graphicx}%
\usepackage{multirow}%
\usepackage{amsmath,amssymb,amsfonts}%
\usepackage{amsthm}%
\usepackage{mathrsfs}%
\usepackage[title]{appendix}%
\usepackage{xcolor}%
\usepackage{textcomp}%
\usepackage{manyfoot}%
\usepackage{booktabs}%
\usepackage{algorithm}%
\usepackage{algorithmicx}%
\usepackage{algpseudocode}%
\usepackage{listings}%

\begin{document}

\title{The Cheap Hours Don't Stay Cheap: Price-Chasing Data Centers Reshape Grids and Markets}

\author*[1]{\fnm{Shijie} \sur{Pan}} \email{shijie.pan@kaust.edu.sa}

\author[1]{\fnm{Zaint A.} \sur{Alexakis}} \email{zaint.alexakis@kaust.edu.sa}

\author[1,2]{\fnm{Charalambos} \sur{Konstantinou}} \email{charalambos.konstantinou@kaust.edu.sa}

\affil[1]{%
  \orgdiv{Computer, Electrical and Mathematical Sciences and Engineering Division},
  \orgname{King Abdullah University of Science and Technology},
  \orgaddress{%
    \city{Thuwal},
    \postcode{23955-6900},
    \country{Saudi Arabia}%
  }%
}
\affil[2]{%
  \orgdiv{Energy Institute},
  \orgname{The University of Texas at Austin},
  \orgaddress{%
    \city{Austin},
    \state{Texas},
    \postcode{78712},
    \country{USA}%
  }%
}

\abstract{Data centers unsettle the paradigm of location-bound electricity demand: computing workloads can migrate across sites and hours, whereas facilities and grid connections remain fixed. This mobility reduces electricity costs but reshapes the spatiotemporal structure of nodal prices. A model-driven framework couples US wholesale market signals with optimal fleet scheduling, showing how their interaction alters prices and other participants' responses. Greater flexibility enlarges savings yet can erode network benefits: in a Texas study, temporal shifting and cross-site allocation each relieve congestion, but their combination leaves total congestion nearly unchanged. At continental scale, computing demand gravitating toward cheaper electricity narrows price differences and compresses the low-price hours storage arbitrage relies on, threatening storage investment recovery. Cost-minimizing scheduling may therefore undermine grid-supporting resources. Comparing job-level and aggregate fleet representations reveals a trade-off between operational visibility and information disclosure, informing requirements and incentives that could turn workload mobility into a lever for system-wide efficiency.}

\maketitle

\section{Introduction}
\label{sec:introduction}

The expansion of AI and cloud computing is increasing both the scale and concentration of electricity demand. The International Energy Agency (IEA) estimates that a typical AI-focused data center consumes as much electricity as approximately 100,000 households~\cite{iea2025energyai}, and projects global data center electricity consumption to rise from approximately 485 TWh in 2025 to 950 TWh in 2030~\cite{iea2026questions}. Accommodating this demand is particularly challenging where grid constraints already limit power delivery, making electricity availability a constraint on further computing expansion~\cite{shehabi20242024,nature2026planning}. At this scale, the timing and location of computing activity become critical to grid operation, since such activity can reshape the spatiotemporal distribution of electricity demand and thereby affect power flows, network operating conditions and, potentially, market opportunities.

\begin{figure*}[t]
\centering
\includegraphics[width=\textwidth]{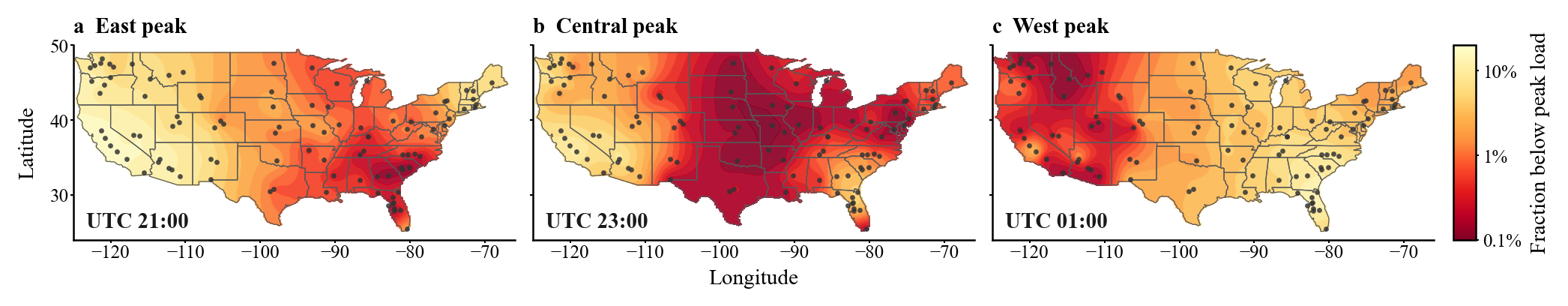}
\caption{\textbf{Regional electricity demand relative to local peaks across the United States.}
Average hourly demand profiles for July 2025 are shown at 21:00, 23:00 and 01:00 UTC, displayed on a logarithmic scale. 
Darker colors indicate demand closer to the local peak.
Dots mark the geographic reference locations, and contours interpolate the calculated values.}
\label{fig:regional_demand_timing}
\end{figure*}

Regional demand patterns provide an intuitive starting point for understanding how the timing and location of electricity consumption matter. Differences in time zones and local demand patterns lead regions to approach their daily demand peaks at different hours (Fig.~\ref{fig:regional_demand_timing}). As one region approaches its evening peak, another may still have lower demand or may already have passed its peak. This variation in demand contributes to differences in electricity prices across locations and hours, alongside generation conditions and network constraints.

Such price differences create an incentive for data center operators to adjust when and where eligible computing workloads are executed. Conventional load-shifting demand response at fixed locations changes the aggregate demand profile by moving consumption toward hours when electricity is less costly to supply. Computing workloads can similarly be deferred or paused and resumed within permitted windows~\cite{crozier2025potential}, while vertical scaling adjusts the resources allocated to running jobs~\cite{pavlenko2024vertically}. Geographically distributed computing, however, offers another response when execution cannot readily be postponed: eligible work can be reassigned to another site within the same hour. Rather than shifting consumption between hours, this response redistributes demand across the network and can leave aggregate consumption largely unchanged. Nevertheless, these temporal and spatial options do not make computing demand freely movable: job arrivals, deadlines and available capacity determine which adjustments are feasible. Within these constraints, experimental work has demonstrated how workload controls can adjust demand while maintaining the tested service requirements~\cite{colangelo2026ai}.

Delivering a feasible workload schedule, however, does not establish its benefit to the grid. The difficulty in assessing price-responsive computing lies in connecting economically attractive workload shifts to the demand changes that can actually be delivered and their effects on the grid. In many US wholesale electricity markets, locational marginal prices (LMPs) reflect the incremental cost of supplying electricity at each node under a given operating state, including the effects of transmission constraints. Large workload shifts can change that state and the prices associated with it. The central question is therefore how the pursuit of lower electricity costs shapes executable demand patterns, how those patterns affect network operating conditions and how the resulting changes reshape the original price opportunities.

Previous research has established how computing workloads can adjust electricity demand under operational constraints~\cite{cao2022assessment,zhou2024hpc} and developed scheduling and market mechanisms to coordinate these adjustments with power-system operation~\cite{zhou2024energy,zhang2022exploring,zhang2020flexibility,zhang2022remunerating,chen2025spatial,li2023computation}. These approaches support cleaner electricity consumption~\cite{radovanovic2022carbon,riepin2025spatio}, grid services~\cite{zhang2021hpc,colangelo2026ai} and network planning~\cite{cao2024managing}. In particular, Kim et al.~\cite{kim2026flexibility} show that predefined operating envelopes can expand feasible data center siting options while largely preserving system-wide mean prices, evaluating candidate sites independently.

These findings motivate a distinct operational question: what happens when operators choose workload allocations across facilities in pursuit of economic returns, rather than follow predefined operating profiles? Workload transfers can introduce load fluctuations and congestion~\cite{zhang2025mitigating}, and expanding scheduling choices need not preserve the network benefits available from more limited responses. Moreover, demand moving toward inexpensive sites and hours can change the prices guiding subsequent allocation and the arbitrage opportunities available to other energy storage and flexible-demand assets, such as batteries, pumped-storage hydropower plants and hydrogen electrolyzers, which typically recover their capital investment through intertemporal price arbitrage. Understanding these consequences requires linking job-constrained fleet scheduling with network operation and price feedback, extending the assessment from the accommodation of individual loads to the system-wide effects of economically driven workload reallocation.

In light of this, the present work links regional electricity price opportunities, job-level workload scheduling and network operation to examine how computing decisions influence demand, grid conditions and market prices. Specifically, it characterizes price differences by magnitude and directional bias, revealing how the spatiotemporal coupling of electricity price incentives shapes opportunities for data center siting and operation. These opportunities are then evaluated through job-level scheduling using production-derived workloads~\cite{duan2026GFS} and a computing-to-power model~\cite{dayarathna2015data}. Results for a Texas case study reveal that substantial electricity cost savings are predominantly accompanied by higher demand peaks and sharper site-level demand ramps. Meanwhile, transmission congestion is redistributed across lines and hours with little change in its overall extent, in contrast to the congestion relief typically expected from conventional demand response. Comparisons of scheduling configurations further show that combining spatial and temporal scheduling freedom can eliminate the congestion relief that either provides alone. Beyond these physical effects, market-driven data center scheduling not only leaves a trace on the network but also alters the market landscape itself. A nationwide analysis shows that computing demand can narrow price spreads and shift the directional bias of regional advantages, reducing storage arbitrage opportunities and potentially weakening capital cost recovery. 
To complement this analysis, aggregate scheduling models are examined in terms of how grid operators can employ them to predict the network effects of workload shifts and the resulting changes in price opportunities, which may substantially affect the operation of connected assets.

\section{Results}
\label{sec:results}

\subsection{Directional bias distinguishes spatial allocation opportunities}
\label{subsec:results_interstate}

The emergence of large, geographically distributed data centers makes it possible to shift substantial electricity demand across regions through workload scheduling. The facilities remain in place, but moving computation allows the associated demand to move between distant locations in response to electricity prices. Regional price relationships therefore influence where demand is attracted and whether that allocation changes over time. Differences in demand timing (Fig.~\ref{fig:regional_demand_timing}) can cause the lower-price location to change over the day, while generation conditions and transmission constraints may sustain an accumulated price advantage at one location even as demand rises and falls.

\begin{figure}[!htbp]
\centering
\includegraphics[width=\textwidth]{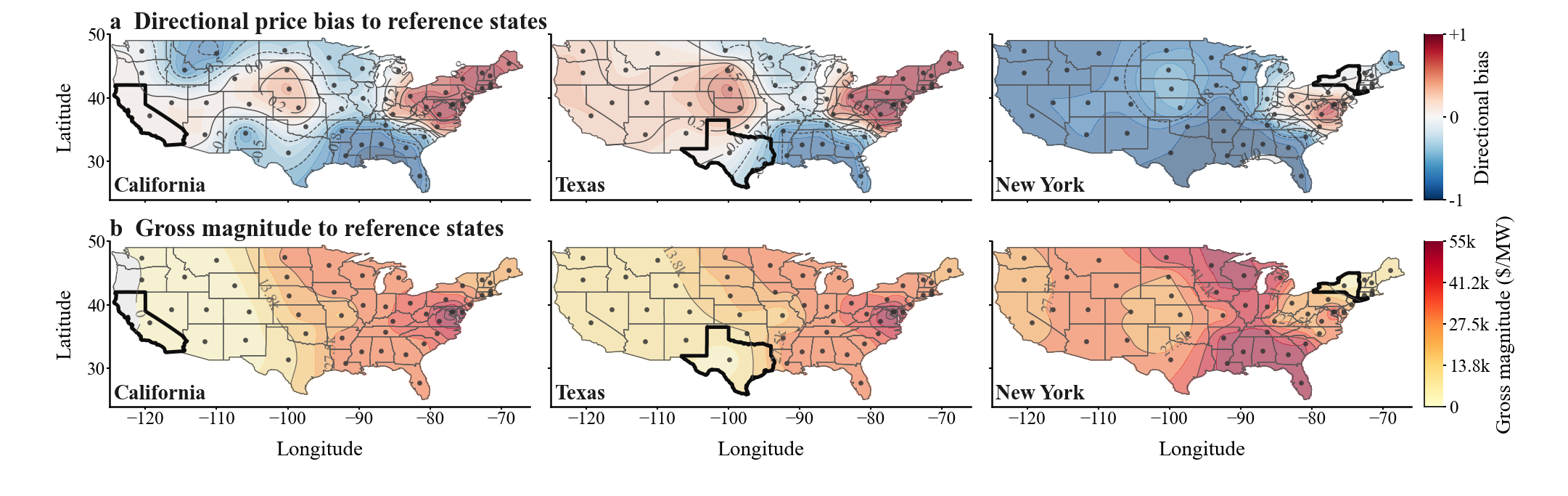}
\caption{\textbf{Magnitude and direction of regional day-ahead electricity-price differences in July 2025.}
Hourly prices from market hubs and regional reference locations are compared over all 744 hours of July.
\textbf{a}, Directional bias relative to the California, Texas and New York reference price series. Bias is calculated as the difference between positive and negative accumulated price spreads divided by their combined magnitude. Positive values indicate that the comparison location is more expensive in accumulated spread magnitude; negative values indicate the reverse. Values near zero indicate balanced positive and negative spread areas.
\textbf{b}, Gross magnitude, calculated by integrating the absolute hourly price difference over the observation window, in \$/MW, that is, the accumulated price difference per MW of demand present in every hour. Dots mark the geographic locations assigned to the price series, contours interpolate the calculated statistics, and thick outlines identify the reference states.}
\label{fig:interstate_lmp}
\end{figure}

The resulting allocation opportunities depend on both the magnitude of price differences and their directional bias. A predominantly one-sided advantage can favor allocating eligible work mainly to one location, whereas substantial advantages in both directions can make adjusting execution locations over time attractive. Figure~\ref{fig:interstate_lmp} jointly quantifies and maps these features for regional day-ahead prices. Their combination distinguishes allocation opportunities that would remain obscured if regional price differences were characterized by magnitude alone.

The California and Texas reference panels show positive directional bias across parts of the Northeast and negative bias across much of the South, with more mixed patterns elsewhere (Fig.~\ref{fig:interstate_lmp}\textbf{a}). Negative bias is more widespread when New York is used as the reference, indicating that its reference series has a higher monthly mean price than many comparison series. Such higher prices may reflect higher marginal generation costs and transmission constraints that limit access to cheaper supply.

The magnitude panels reveal a complementary pattern (Fig.~\ref{fig:interstate_lmp}\textbf{b}). California and Texas show relatively small spreads across parts of the West and larger spreads toward the eastern United States. Their similar patterns are consistent with the comparatively modest mean absolute difference of 13.6 \$/MWh between the two reference series. Using New York as the reference reduces the magnitude of some northeastern comparisons while increasing differences across parts of the South. In these southern comparisons, larger spreads coincide with negative directional bias, indicating a pronounced accumulated price advantage relative to the New York reference.

\begin{table}[!htbp]
\centering
\caption{\textbf{Representative day-ahead price differences relative to Texas in July and January 2025.} Each month contains 744 paired hourly observations.}
\label{tab:spread_examples}
\small
\begin{tabular}{lrrrr}
\toprule
 & \multicolumn{2}{c}{July}
 & \multicolumn{2}{c}{January} \\
\cmidrule(lr){2-3}\cmidrule(lr){4-5}
Comparison
& \shortstack{Directional\\bias}
& \shortstack{Mean absolute\\spread (\$/MWh)}
& \shortstack{Directional\\bias}
& \shortstack{Mean absolute\\spread (\$/MWh)} \\
\midrule
California & 0.30 & 13.6 & $-$0.75 &  61.9 \\
Michigan   & 0.03 & 40.1 & $-$0.88 &  78.5 \\
New York   & 0.90 & 38.2 &  0.70   & 153.7 \\
\bottomrule
\end{tabular}
\end{table}

Similar price differences can offer markedly different opportunities for allocating computing workloads across data center sites. In July, Michigan and New York exhibit comparable mean absolute price differences from Texas, at 40.1 and 38.2~\$/MWh, respectively, yet their directional biases are 0.03 and 0.90 (Table~\ref{tab:spread_examples}). For Michigan, the near-zero bias indicates that the accumulated price advantage is almost evenly split between the two sites; combined with the large mean absolute spread, this signals substantial bidirectional arbitrage potential, which can be captured only by identifying the hours favorable to each location. For New York, the bias of 0.90 indicates that the accumulated price difference overwhelmingly favors Texas, so workloads would migrate almost exclusively toward Texas; a New York site would therefore seldom attract shifted demand, offering little economic justification for pairing it with Texas for spatial arbitrage. These opportunities also change seasonally: January reverses the accumulated advantage from Texas to California, while Michigan's nearly balanced July comparison shifts toward a pronounced advantage for Michigan (Table~\ref{tab:spread_examples}; Supplementary Fig.~1).

\subsection{Electricity savings accompany higher demand peaks}
\label{subsec:results_jobflex}

The price differences identified above create opportunities for workload scheduling, but exploiting them depends on job arrivals, execution requirements and available site capacity. To examine how cost-minimizing schedules translate into electricity consumption and expenditure, a case study is conducted across 11 modeled Texas sites grouped into six data center operator portfolios (Supplementary Fig.~2), using projected capacities for 2030~\cite{epochAI} and production-derived workloads~\cite{duan2026GFS}. These outcomes are evaluated by comparing price-responsive scheduling against a price-independent reference. The price-responsive case combines the temporal and spatial adjustments described above with changes in the computing resources allocated to individual jobs. Workloads may migrate only among sites belonging to the same operator, and jobs may be terminated before completion, subject to a service-level agreement (SLA) requiring each operator to deliver at least 95\% of its total requested computing work. Eligible jobs may also run up to 24 hours beyond the original 168-hour workload window. The reference case, in contrast, executes jobs in arrival order subject to available site capacity, delivering all requested work within the original window. Texas schedules respond to observed ERCOT day-ahead prices and are then evaluated in a network model without updating those prices; this isolates the physical effect of the scheduling response, while price feedback is examined separately in the continental analysis.

\begin{figure}[!htbp]
\centering
\includegraphics[width=\textwidth,height=0.64\textheight,keepaspectratio]{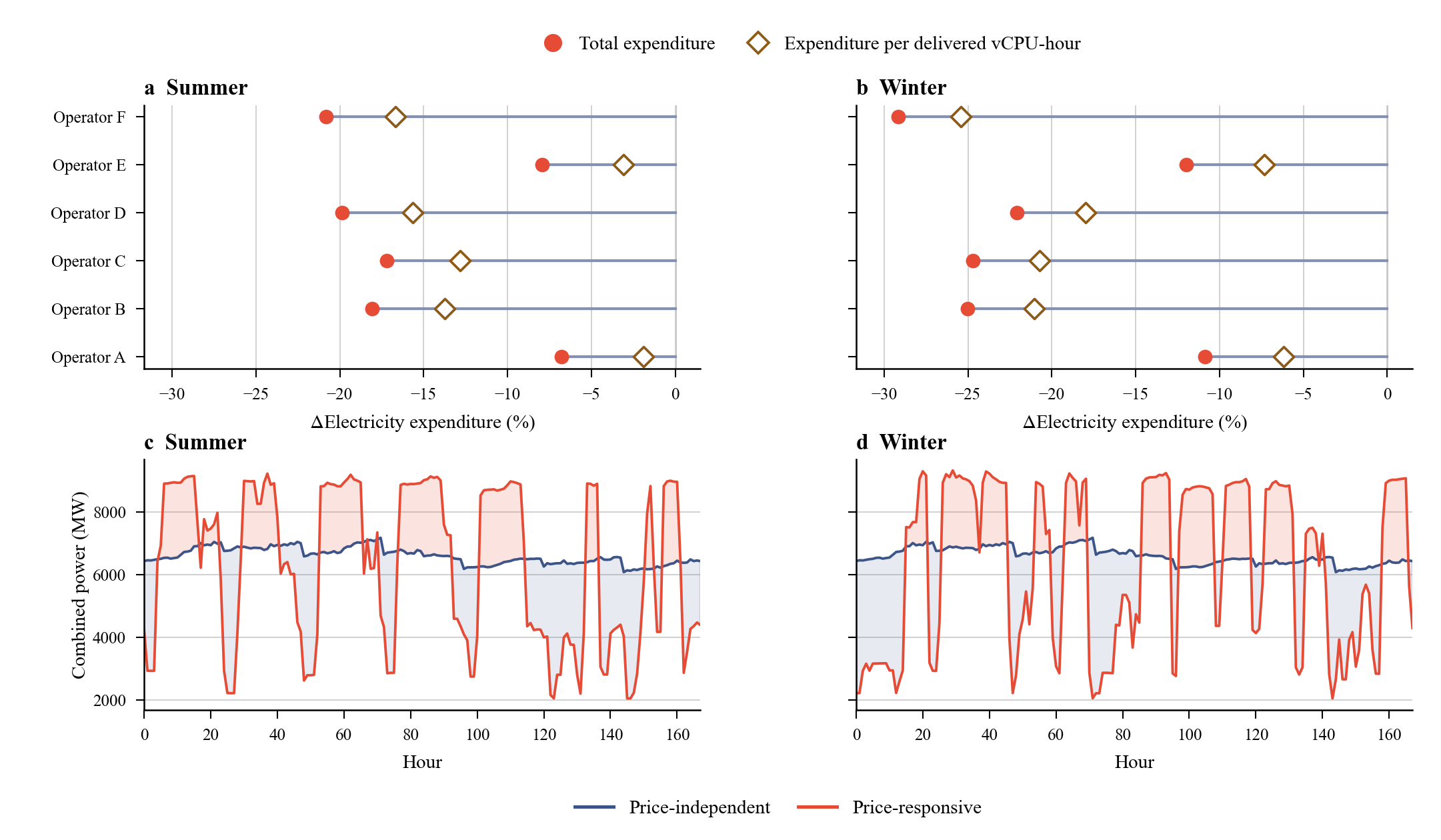}
\caption{\textbf{Electricity savings and demand concentration under price-responsive scheduling.}
\textbf{a,b}, Changes in electricity expenditure for each modeled data center operator relative to price-independent scheduling under summer and winter prices, respectively.
Circles show the change in total electricity expenditure, and diamonds show the change after normalizing expenditure by delivered computing work in vCPU-hours.
Expenditure includes dynamic power costs over the 192-hour scheduling horizon and idle power costs over the original 168-hour window.
Price-responsive scheduling requires at least 95\% delivery of each data center operator's requested computing work and permits eligible execution during an additional 24-hour window; the price-independent reference delivers all represented requested work.
\textbf{c,d}, Combined power demand across all 11 modeled sites during the original 168-hour summer and winter windows.
Blue and red identify price-independent and price-responsive scheduling, respectively.}
\label{fig:job_flexibility}
\end{figure}

Electricity expenditure falls across all six operator groups, although the size of the reduction differs substantially: 6.8-20.8\% under summer prices and 10.9-29.2\% under winter prices (Fig.~\ref{fig:job_flexibility}\textbf{a,b}). Percentage reductions are generally larger under winter prices. Operators A and E each comprise one modeled site, so their savings arise from temporal adjustments within that site. The remaining operators include multiple sites, allowing eligible workloads to be rescheduled across both locations and hours. After normalizing expenditure by delivered computing work, reductions remain at 1.9-16.7\% in summer and 6.2-25.4\% in winter, indicating that lower expenditure is not explained solely by reduced work delivery (Fig.~\ref{fig:job_flexibility}\textbf{a,b}). Savings also persist when all represented work must be completed within the original job windows, with total expenditure reductions of 13.3\% in summer and 18.4\% in winter relative to the corresponding price-independent reference (Supplementary Fig.~3). 

The savings coincide with higher demand peaks and deeper troughs (Fig.~\ref{fig:job_flexibility}\textbf{c,d}). Under price-independent scheduling, combined demand across all modeled sites remains between approximately 6.08 and 7.18~GW during the evaluated week. Under price-responsive scheduling, combined demand peaks at 9.23~GW under summer prices and 9.33~GW under winter prices, respectively 28.6\% and 30.0\% above the price-independent peak. These demand changes arise under an idealized configuration without explicit site-level ramp limits or reallocation penalties.

Elevated demand persists across the week: combined demand exceeds 8~GW in 69 summer hours and 73 winter hours, whereas the price-independent reference remains below that level throughout. The changes also extend to individual sites. Without explicit site-level ramp limits, the maximum absolute hourly change in demand at a single site increases from approximately 0.19~GW under price-independent scheduling to 1.60~GW under price-responsive scheduling in both summer and winter (Supplementary Fig.~4). The combined demand curves do not fully capture these local changes: similar changes in total demand can arise from different distributions of consumption across sites. Evaluating the network response therefore requires retaining the timing and location of individual site demands. How the grid accommodates these changes determines whether the electricity savings are accompanied by improved network conditions.

\subsection{More scheduling options do not guarantee greater congestion relief}
\label{subsec:results_congestion}

Responding to electricity prices by allocating computing workloads to cheaper locations and hours may be expected to benefit grid operation. However, this response also redistributes substantial demand across the network, with the higher peaks and sharper site-level changes identified above.

To evaluate the network effects of these schedules, the hourly demand profiles of the 11 data center sites are incorporated into an equivalent transmission network model of the Texas power system, thereby linking the cost savings and demand changes identified above to transmission loading. Three cases are compared under summer and winter conditions: network operation without explicit data center demand, with price-independent scheduling, and with price-responsive scheduling. Both seasonal evaluations share the network configuration and background demand of the winter dataset, with the summer evaluation substituting July solar and onshore wind availability profiles.

\begin{figure}[!htbp]
\centering
\includegraphics[width=\textwidth,height=0.66\textheight,keepaspectratio]{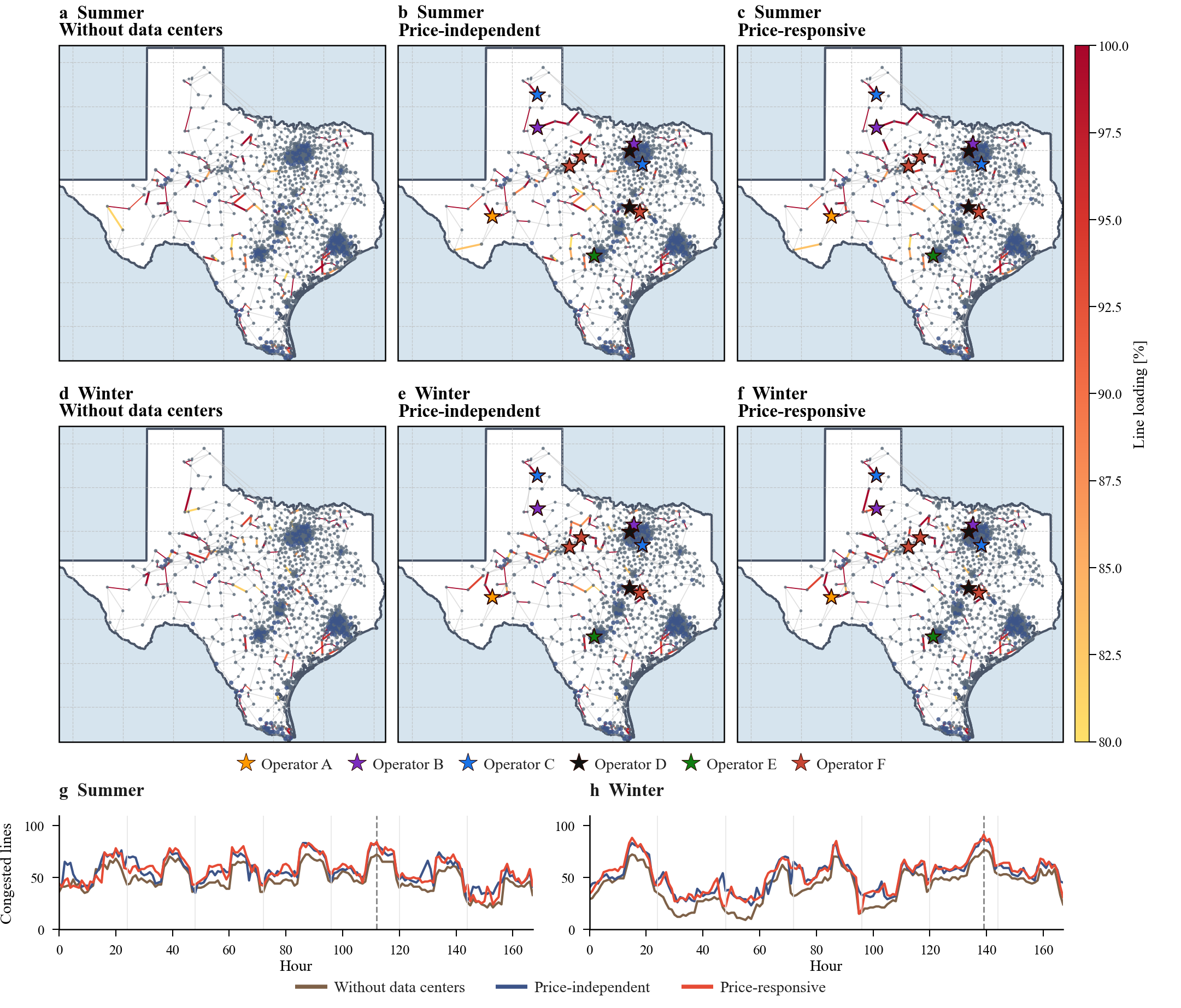}
\caption{\textbf{Spatial and temporal redistribution of transmission loading under data center scheduling.}
\textbf{a-c}, Summer transmission line loading without the explicitly modeled data center demand, under price-independent scheduling, and under price-responsive scheduling, respectively.
\textbf{d-f}, Corresponding winter results.
Maps show hour 112 in summer and hour 139 in winter, when the number of congested lines under price-responsive scheduling is greatest.
Within each season, all three maps show the same hour.
Line colors indicate loading between 80\% and 100\% of the line rating.
Stars identify modeled sites, with colors distinguishing data center operators.
\textbf{g,h}, Hourly counts of congested lines during the 168-hour summer and winter windows.
Vertical dashed lines identify the hours shown in the maps.}
\label{fig:congestion_relocation}
\end{figure}

Despite the pronounced changes in computing demand, the total number of congested line-hours changes little. Relative to price-independent scheduling, it increases by 0.1\% under summer conditions and 1.2\% under winter conditions. The maps and hourly profiles reveal a more varied response beneath these small net changes, with relief in some locations and hours accompanied by increases elsewhere (Fig.~\ref{fig:congestion_relocation}).

Across both seasons, the changes extend along western and central transmission paths beyond the lines adjacent to data center sites (Fig.~\ref{fig:congestion_relocation}\textbf{a-f}). The scheduling curves cross repeatedly over the week, indicating that price-responsive demand improves network conditions at some hours and worsens them at others (Fig.~\ref{fig:congestion_relocation}\textbf{g,h}). Relative to price-independent scheduling, price-responsive scheduling introduces 1,699 congested line-hours while relieving 1,689 in summer; the corresponding winter values are 1,311 and 1,209. Net increases of only 10 and 102 line-hours thus reflect congestion relief being offset by new congestion elsewhere. Under explicit site-level ramp limits, large volumes of newly congested and relieved line-hours remain, while the net change stays small relative to them (Supplementary Fig.~5), suggesting that restricting the speed of demand changes alone does not ensure congestion relief.

The higher demand peaks identified above do not coincide with the most widespread congestion. Combined data center demand reaches its maximum at hour 37 in summer and hour 29 in winter (Fig.~\ref{fig:job_flexibility}\textbf{c,d}), whereas the number of congested lines under price-responsive scheduling peaks at hours 112 and 139, respectively. This separation shows why the combined computing demand curve alone cannot identify when congestion is most widespread: the network must accommodate the geographic distribution of computing demand together with background demand and available generation.

\begin{figure}[!htbp]
\centering
\includegraphics[width=\textwidth,height=0.66\textheight,keepaspectratio]{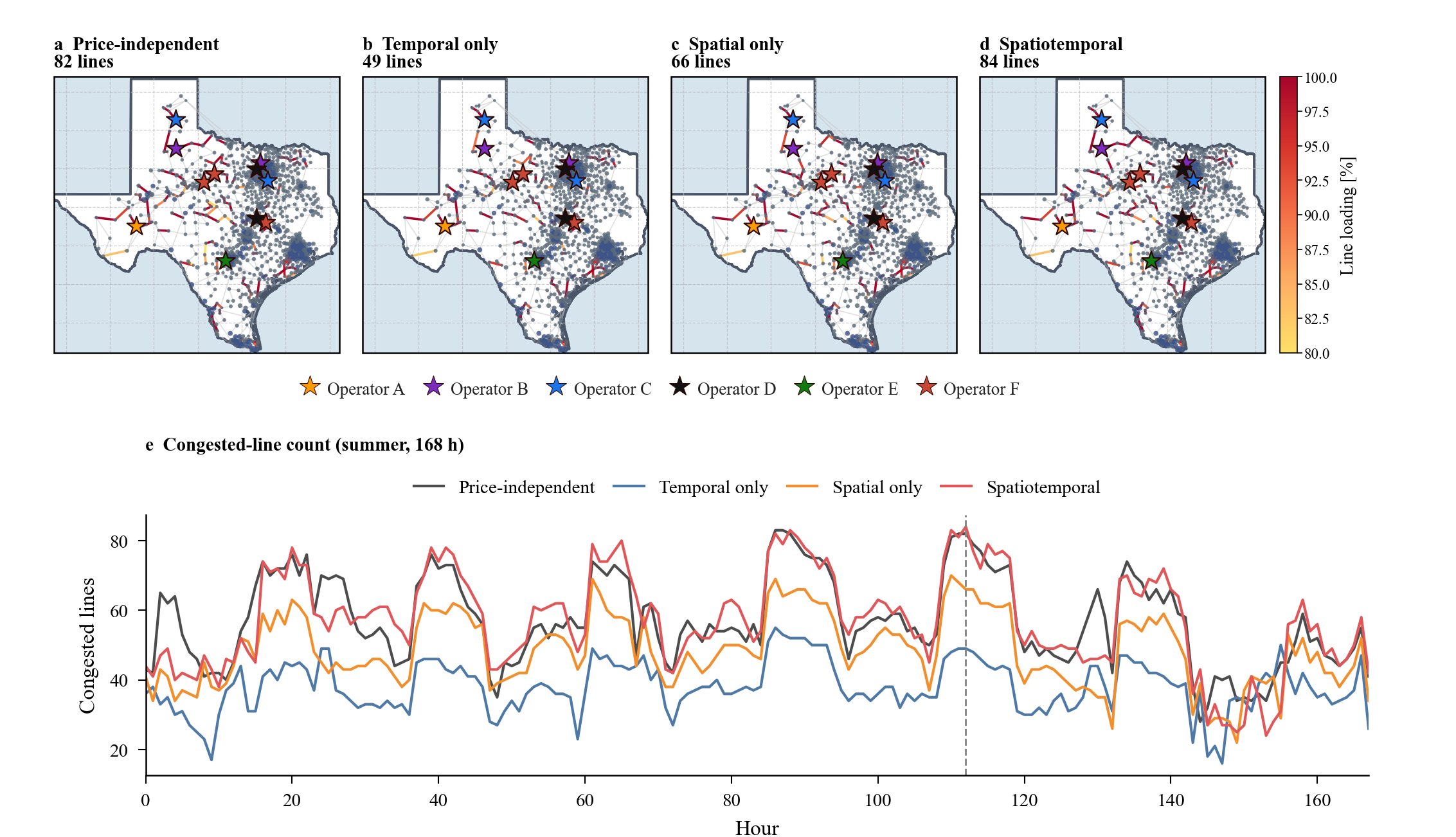}
\caption{\textbf{Transmission loading under different workload scheduling configurations.}
\textbf{a-d}, Summer transmission line loading under price-independent scheduling, single-site scheduling with extended execution windows (temporal only), cross-site reallocation within the original windows (spatial only), and their combination (spatiotemporal), respectively. The temporal-only configuration permits initial site selection but prevents transfers during execution; the spatial-only configuration permits cross-site reallocation and resource adjustments only within the original windows. 
\textbf{e}, Hourly counts of congested lines over the original 168-hour window.
The vertical dashed line identifies the hour shown in the maps.}
\label{fig:four_mode_congestion}
\end{figure}

To further examine how temporal adjustments and cross-site reallocation shape the network response, four illustrative configurations under summer conditions are compared: price-independent scheduling, temporal-only scheduling with extended execution windows (each job may be assigned to any site of its operator but cannot move once started), spatial-only reallocation within the original windows, and their combination (Fig.~\ref{fig:four_mode_congestion}). Relative to price-independent scheduling, total electricity expenditure falls by 10.4\%, 13.3\% and 17.6\% under these three price-responsive configurations, respectively. Congested line-hours, however, decrease by 33.5\% and 15.5\% under temporal-only and spatial-only scheduling, respectively, and  increase marginally by 0.1\% when the two are combined.

The temporal-only configuration resembles conventional temporal demand response by allowing execution to be deferred toward inexpensive hours. Its congestion relief is consistent with the potential for temporal demand response to improve network conditions (Supplementary Fig.~6). Allowing cross-site reallocation instead opens another route to lower expenditure: operators can exploit price differences between buses without relying solely on postponing work. This option is particularly relevant when deadlines limit deferral. Although such reallocation can leave total demand at congested hours largely unchanged, it can relieve individual transmission constraints by moving consumption away from locations that load constrained lines more heavily.

Yet combining these options shows that greater scheduling freedom can increase private savings while eliminating the congestion relief achieved through either option alone. Expanding scheduling choices therefore need not preserve their individual network benefits, as demand may become concentrated at inexpensive sites and hours. Such concentration may also change the prices that motivated it, raising a further question: how does workload relocation reshape the spatial price differences from which it seeks to benefit?

\subsection{Computing demand reshapes prices and storage arbitrage opportunities}
\label{subsec:results_endogenous}

As computing demand grows and moves across sites and hours, it reshapes regional demand patterns (Supplementary Fig.~7) and can change the prices guiding subsequent workload allocation. To examine this feedback, a continental network analysis is performed in which workload schedules and dispatch-derived marginal prices are iteratively updated to capture how each responds to changes in the other (Supplementary Fig.~8). The price-feedback procedure is detailed in Methods (``Continental marginal prices and price feedback'').

\begin{figure}[!htbp]
\centering
\includegraphics[width=\textwidth]{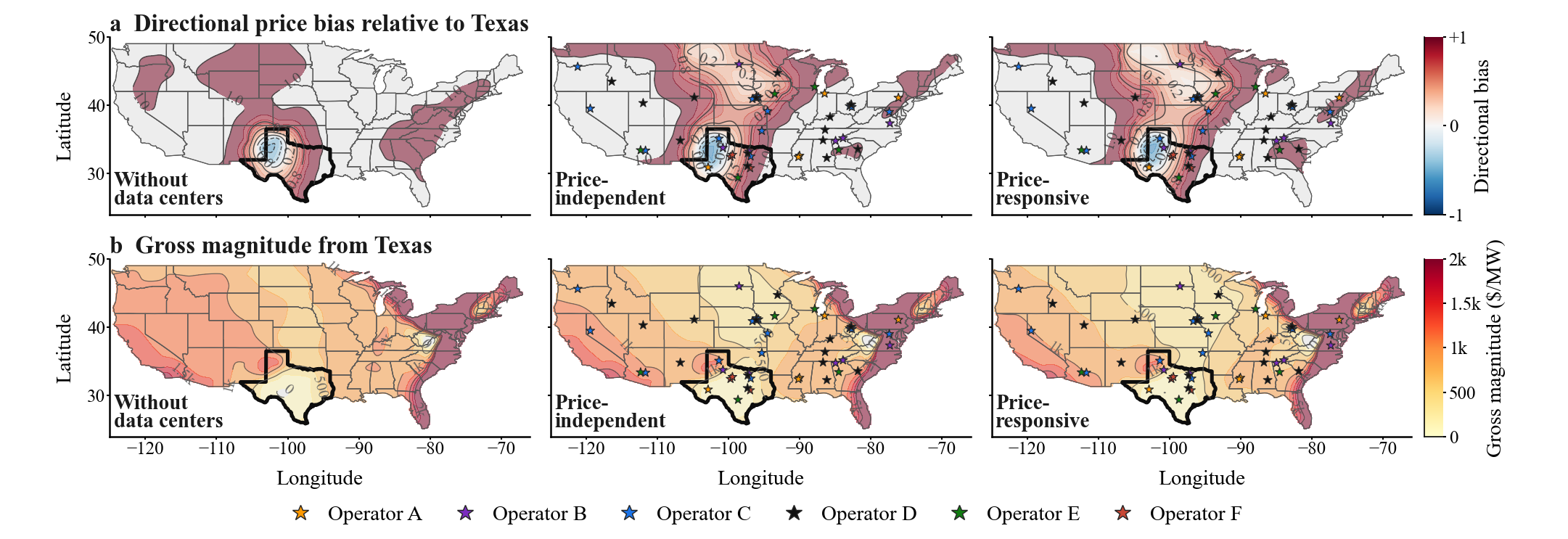}
\caption{\textbf{Spatial price differences after computing demand responds to updated electricity prices.}
Marginal prices are evaluated over one summer week without the explicitly modeled data center demand, under price-independent scheduling, and under price-responsive scheduling after price updates.
\textbf{a}, Directional bias relative to the Texas reference node.
\textbf{b}, Gross magnitude, calculated by integrating the absolute hourly price difference relative to the same reference over the 168-hour window, in \$/MW.
Contours interpolate bus-level statistics, stars mark modeled data center locations, and the thick outline identifies Texas.}
\label{fig:endogenous_lmp}
\end{figure}

Price-responsive scheduling narrows average spatial price differences, but the adjustment also changes which locations retain an accumulated advantage (Fig.~\ref{fig:endogenous_lmp}). After price updates, the mean absolute price difference between modeled data center sites is 5.3\% lower than under price-independent scheduling, and the mean gross magnitude relative to Texas across all buses is 2.9\% lower. The directional response reveals changes that these averages conceal. At one northern node, directional bias shifts from $+$0.98 without explicitly modeled data center demand to $-$0.07 under price-independent scheduling and $-$0.22 under price-responsive scheduling, reversing the accumulated advantage from Texas to the comparison location. Across nodes, directional-pattern correlations relative to the scenario without modeled data centers are 0.80 under price-independent and 0.76 under price-responsive scheduling, suggesting that added computing demand accounts for much of the departure, while price responsiveness introduces a further change. Substantial regional contrasts remain, with smaller gross magnitudes across parts of the interior network and larger differences in the eastern United States (Fig.~\ref{fig:endogenous_lmp}\textbf{b}). The corresponding winter analysis shows that the magnitude and spatial distribution of these price changes are season dependent (Supplementary Fig.~9). 

\begin{figure}[!htbp]
\centering
\includegraphics[width=\textwidth]{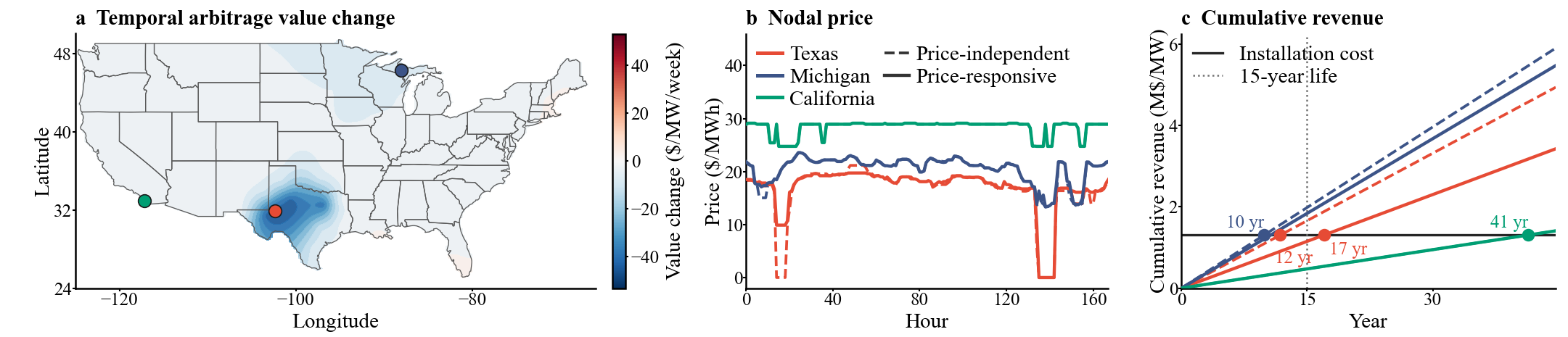}
\caption{\textbf{Data center scheduling changes the temporal arbitrage opportunities and long-term economics of battery storage.}
The comparison in \textbf{a,b} uses the summer marginal prices from Fig.~\ref{fig:endogenous_lmp}.
A 1~MW/4~MWh battery with 90\% round-trip efficiency is evaluated separately at each bus as a price taker.
\textbf{a}, Change in potential battery energy-arbitrage value, calculated as price-responsive minus price-independent and normalized by battery power capacity, in \$/MW/week.
Negative values indicate reduced arbitrage value.
Contours interpolate bus-level results; colored markers identify the representative nodes shown in \textbf{b}.
\textbf{b}, Hourly marginal prices at selected nodes in Texas, Michigan and California under price-independent (dashed) and price-responsive (solid) computing schedules.
\textbf{c}, Illustrative cumulative energy-arbitrage revenue at the selected nodes.}
\label{fig:battery_arbitrage}
\end{figure}

However, the economic effects of this price response extend beyond workload allocation by data center operators. These price changes also affect storage, whose arbitrage revenue depends on temporal price differences. Large-scale storage itself is known to compress these spreads~\cite{sioshansi2009estimating,lamp2022large}; here, the effect of computing demand is evaluated by optimizing battery arbitrage separately at each network bus, using the same battery specifications under price-independent and price-responsive computing demand. The reduction in potential arbitrage value is most pronounced around Texas (Fig.~\ref{fig:battery_arbitrage}\textbf{a}). At the representative Texas node, the weekly arbitrage value falls from 154.4 to 102.7~\$/MW, a reduction of 33.5\%. The corresponding reduction at the selected Michigan node is 12.7\%, whereas the value at the selected California node changes little. The hourly price profiles illustrate the mechanism behind this spatial heterogeneity (Fig.~\ref{fig:battery_arbitrage}\textbf{b}). At the Texas node, prices in an early near-zero interval under price-independent computing rise to approximately 10~\$/MWh under price-responsive scheduling, compressing the price spread available for battery charging. Although another near-zero-price interval remains later in the week, the loss of part of the earlier low-price opportunity reduces the battery's weekly arbitrage value. The concentration of 11 modeled data center sites in Texas may strengthen this interaction: shifting substantial computing demand toward inexpensive hours raises prices precisely when storage would otherwise prefer to charge. As demand grows, its composition also matters: ordinary-load growth through low-price siting and the expansion of price-responsive computing produce different changes in storage arbitrage opportunities, even when they add the same electricity consumption (Supplementary Fig.~10).

Such compression of price spreads may also affect the long-term economics of storage. To assess this effect, annualized arbitrage revenues estimated from observed prices are scaled by the modeled change in mean daily price ranges. Because the annual baseline comes from observed prices rather than the single modeled week, and the scaling uses daily price ranges rather than the weekly arbitrage value, the payback periods do not follow directly from the weekly values in Fig.~\ref{fig:battery_arbitrage}\textbf{a}. Under this illustrative adjustment, the payback period increases from approximately 12 to 17 years at the selected Texas node and from approximately 10 to 11 years at the Michigan node, while remaining approximately 41 years at the California node  (Fig.~\ref{fig:battery_arbitrage}\textbf{c}). In Texas, this estimate exceeds the assumed 15-year battery lifetime; if the modeled weekly compression were representative of longer-term conditions, arbitrage alone would no longer recover the investment. A complementary analysis of pumped-storage hydropower illustrates the same interaction in another storage technology (Supplementary Fig.~11).

\subsection{Information requirements for effective grid-data center coordination}
\label{subsec:results_agg}

Although the preceding investigations show that workload scheduling can have significant implications for power-system operation and electricity markets, an important question remains: how can grid operators characterize and anticipate these effects? Grid operators rely on simulation studies in which large and complex loads are represented through equivalent or surrogate models. It remains unclear how much detail these models must retain to capture the effects identified above while remaining tractable for system-level studies. To examine this question, three aggregate representations based on energy requirements and power limits are considered, with reference quantities obtained from the job-level solution to provide a deliberately favorable comparison. AGG Site matches each site's total electricity consumption while allowing it to shift across hours; AGG Operator matches only the total across each operator's sites, allowing consumption to shift across both sites and hours; AGG Continuous retains the same site-level power bounds but imposes no total energy requirement. Their construction and the model formulations are detailed in Methods (``Aggregate scheduling models'').

\begin{figure}[!htbp]
\centering
\includegraphics[width=\textwidth]{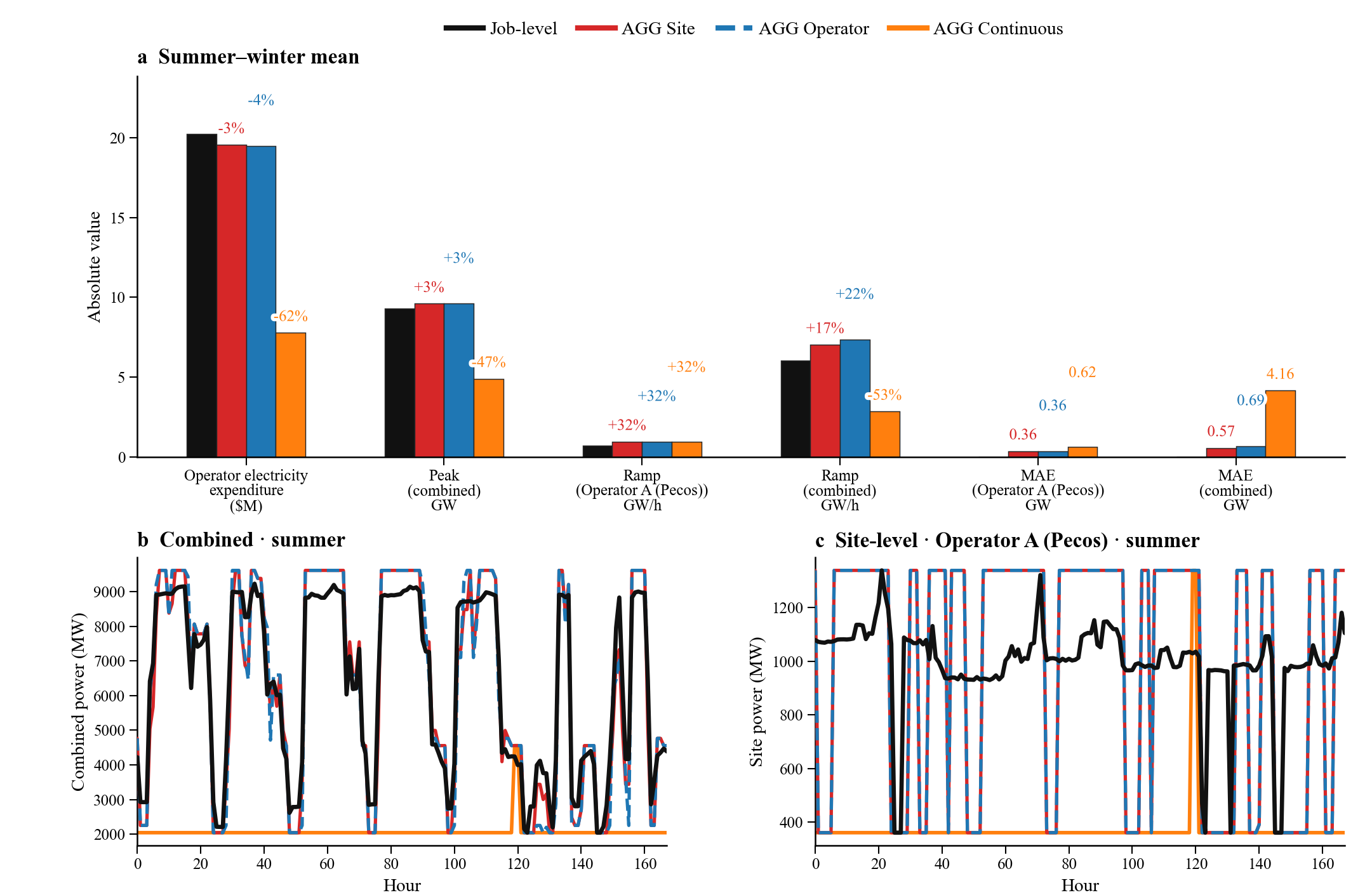}
\caption{\textbf{Similar electricity expenditure under energy-matched aggregate and job-level scheduling accompanies different demand profiles.}
\textbf{a}, Electricity expenditure, combined peak demand, maximum absolute hourly demand changes at the site of operator A and across all sites, and MAE relative to the corresponding job-level demand profiles.
All metrics are evaluated over the same 168-hour window. Bars show arithmetic means of the summer and winter metrics. Percentage labels indicate changes relative to job-level scheduling; MAE labels show absolute errors in GW.
\textbf{b,c}, Combined demand across all modeled sites and demand at the site of operator A, respectively, during the summer evaluation week.
The operator A profiles illustrate differences in site-level demand that are less apparent in the combined profiles.
AGG Site matches the job-level electricity consumption at each site, whereas AGG Operator matches total consumption across each data center operator's sites.
Both impose power bounds derived from the job-level demand profiles.
AGG Continuous retains these bounds but imposes no energy requirement.}
\label{fig:agg_error}
\end{figure}

The two energy-matched models closely reproduce electricity expenditure and the broad timing of combined demand. Averaged across summer and winter, expenditure is approximately 3\% lower under AGG Site and 4\% lower under AGG Operator than under job-level scheduling, while both retain the main high- and low-demand periods (Fig.~\ref{fig:agg_error}\textbf{a,b}). AGG Continuous instead remains near the lower power limit for much of the week and consumes substantially less electricity. Its lower expenditure therefore cannot be interpreted as a saving from scheduling an equivalent amount of work.

Agreement in the broad demand pattern becomes less complete at individual hours. The energy-matched schedules produce higher, flatter demand plateaus around hours 55-65 and 80-90, while the job-level schedule retains variations within these periods. Averaged across summer and winter, combined peak demand increases from 9.28~GW in the job-level schedules to approximately 9.61~GW in either energy-matched representation. 

The differences become more pronounced at individual sites. At the site of data center operator A, the energy-matched schedules repeatedly move between the lower and upper power bounds, whereas the job-level schedule spends much of the week at intermediate demand levels (Fig.~\ref{fig:agg_error}\textbf{c}). Averaged across summer and winter, the maximum absolute hourly demand change at this site is approximately 32\% higher in both energy-matched representations. The mean absolute error (MAE) relative to the job-level profile is approximately 0.36~GW under both energy-matched representations at this site. For combined demand, the corresponding errors are 0.57~GW under AGG Site and 0.69~GW under AGG Operator (Fig.~\ref{fig:agg_error}\textbf{a}).

The comparison shows that energy requirements and power limits closely reproduce electricity expenditure and the broad temporal pattern of combined demand, but do not fully preserve the hourly demand profiles at individual sites, even when parameterized using reference values obtained from the job-level solution. This exposes a trade-off in the representations examined here: reducing job-level information preserves broad economic outcomes at the expense of fidelity in local demand profiles. 
\section{Discussion}
\label{sec:discussion}

Data centers now act on both the physical and economic layers of the power system, reshaping power flows as well as the price signals intended to coordinate them. At sufficiently large scale, workload allocation can alter transmission conditions and, in turn, the electricity prices that guide subsequent scheduling decisions, creating a physical and economic feedback loop that operates on operational timescales. Prior work has quantified the flexibility of computing workloads, designed mechanisms to coordinate or remunerate it, and assessed its value for emissions reduction, siting and interconnection~\cite{cao2022assessment,zhang2020flexibility,radovanovic2022carbon,kim2026flexibility}. This paper instead examines how the geographical distribution of data center fleets interacts with nodal prices, and how the resulting workload shifts feed back into network operation and the price opportunities available to other resources. The incentive to shift workloads is shaped not by nodal prices in isolation but by their spatial and temporal coupling: spatial shifting pays only when the price difference between two sites is large and favors the destination site during the hours in which work can move. Where these conditions hold, workload relocation reduces electricity costs, and the resulting demand redistribution propagates into the physical operation of the grid. 

Yet New York continues to attract substantial data center investment despite the limited spatial arbitrage value it offers relative to lower-cost locations such as Texas (Table~\ref{tab:spread_examples}). Nearly 12~GW of data center load requests had entered the NYISO interconnection queue by mid-2026, prompting a statewide pause on permits for new facilities above 50~MW~\cite{ny_eo62}. This concentration reflects the proximity requirements of the computation itself: as a leading global financial center, New York hosts trading and financial services that must execute with minimal latency to respond to rapidly evolving market conditions. Where facilities are built, and hence which price couplings a fleet can exploit, is thus shaped jointly by electricity, computing and specific service markets, a dimension beyond the operational scope examined here.

Even favorable prices do not guarantee an executable response. Unlike a battery's power and energy ratings, installed computing capacity alone does not define the demand available for rescheduling: job arrivals, deadlines and service commitments determine how that capacity can be used across sites and hours. Consequently, anticipating the response and its grid impacts requires considering price opportunities together with the workload state. By accounting for this interaction, the analysis reveals that price-driven workload scheduling can produce sharper demand transitions and greater spatial concentration, intensifying congestion in parts of the network rather than systematically relieving it. The origin of this behavior lies in what makes computing demand fundamentally different from conventional forms of grid flexibility.

Demand-response schemes for distributed resources and flexible loads are designed to redirect consumption toward lower-priced periods, much as price-driven data center scheduling does. The underlying mechanism is therefore not new: price-driven computing responds to many of the same economic opportunities as storage, whose large-scale arbitrage is already known to compress price differences~\cite{sioshansi2009estimating,lamp2022large}. Their operating constraints nevertheless differ: storage shifts energy within finite capacity and state-of-charge limits, whereas computing reallocates consumption among existing facilities subject to site capacity and incoming workload requirements. Continuing job arrivals can renew opportunities for workload reallocation throughout the scheduling horizon, allowing large fleets to exert sustained pressure on price differences when suitable work is available. The resulting competition for inexpensive electricity makes storage economics a moving target. Improvements in efficiency, lifetime and capital cost may strengthen project economics, yet their value depends on how the surrounding price environment evolves as computing demand expands. Assessments of storage investment should therefore consider technological progress alongside changes in arbitrage opportunities created by other responsive loads.

Several modeling choices bound these results. Each season is represented by a single week, and the summer Texas evaluation retains winter background demand with July renewable availability. Network effects are evaluated with linear optimal power flow, without reactive power, contingency security or dynamic stability. Data center operators are modeled as price takers, and the main price-responsive cases impose no migration, interruption or ramping costs, so their demand changes represent an upper-bound response. Workloads are drawn from a relatively responsive GPU trace, converted to CPU-equivalent requests and valued at spot computing prices. The price-independent reference schedules jobs first-come, first-served without backfilling, which affects the magnitude of the reported peak increases. Other workload mixes, service requirements and reference schedulers would change the magnitudes reported here, although the mechanism, in which demand concentrates when and where prices are low and thereby alters those prices, is expected to persist wherever large price-responsive demand is available.

This emerging reality also changes what coordination between data centers and power systems must accomplish. Data center operators respond to electricity prices under internal workload constraints, whereas the grid ultimately experiences the demand profiles produced by those decisions. As data center capacity grows, effective coordination increasingly depends on communicating feasible demand adjustments without exposing sensitive job-level information. Aggregate representations offer one means of describing how consumption can vary across individual sites, provided that they retain the characteristics needed for network assessment. Forecasting workload arrivals and execution requirements could complement these representations and reduce reliance on direct disclosure, although uncertainty limits how accurately future demand adjustments can be anticipated. The results therefore expose a fundamental dilemma in grid and data center coordination: greater operational visibility can improve system-level coordination and support more sustainable operation, but often at the cost of increased disclosure of sensitive information, while stronger privacy protection reduces the fidelity with which grid impacts can be anticipated. Neither objective can be pursued in isolation as computing demand continues to expand, making privacy-preserving forms of coordination increasingly necessary.

Grid operators can also complement the information exchanged with data centers through operating requirements and economic signals that account for the scale and volatility of their response. Interconnection and operating requirements for large loads increasingly address load profiles, dynamic behavior, telemetry and operating limits~\cite{ERCOTLargeLoad}. Site-level ramp limits can moderate abrupt demand changes, yet substantial congestion redistribution persists even under stricter limits (Supplementary Fig.~5). Such limits constrain how quickly demand changes without ensuring that its allocation improves network conditions. Economic signals must therefore also account for how large workload responses alter the conditions underlying those signals. Efforts to address congestion risks from price-responsive storage, including the AgNes network tariff reform~\cite{bundesnetzagentur2026agnes}, offer a relevant precedent for considering network consequences in incentive design. As data centers become a structural component of electricity demand, the challenge is therefore not to suppress their flexibility, but to govern it so that the same economic signals that enable efficient computing also preserve the reliability, efficiency and welfare of the grid they increasingly reshape. 
\section{Methods}
\label{sec:methods}

\subsection{Experimental design}

The analysis links observed electricity prices, job-level scheduling and power system models. Texas evaluations assess demand and congestion, continental simulations examine price feedback and battery arbitrage, and aggregate comparisons evaluate the demand information retained without individual jobs.

The Texas analysis includes 11 sites belonging to six data center operator groups. The continental analysis includes 48 sites assigned to 33 buses of a 134-bus network. Each data center operator schedules its own workloads among its own sites; workloads are not exchanged between data center operators. Operators are labeled A-F; each label denotes a modeled portfolio constructed from public facility locations and capacities. The workload traces do not represent the proprietary computing activity of any company. 

The original workload window, $\mathcal{T}_0$, spans 168 hours. Eligible jobs may execute during an additional 24 hours, giving a full scheduling horizon $\mathcal{H}$ of 192 hours. Demand profiles and network evaluations use $\mathcal{T}_0$. Job-level electricity expenditure includes dynamic consumption over $\mathcal{H}$ and idle consumption over $\mathcal{T}_0$, whereas aggregate comparisons use $\mathcal{T}_0$ for both.

\subsection{Data sources and workload preparation}

Regional electricity demand is obtained from the Public Utility Data Liberation (PUDL) project~\cite{pudl_data}, using estimated state-level demand derived from FERC Form-714~\cite{ferc714} and balancing-authority demand from EIA Form-930~\cite{eia930}. January and July 2025 are used to construct average hourly demand profiles. The summer maps include 101 geographic reference locations after filtering for data availability. State and balancing-authority records can overlap geographically. They are used to describe local demand patterns and are not summed to estimate national electricity consumption.

Observed hourly day-ahead electricity prices are accessed through GridStatus~\cite{gridstatus_data}. Texas scheduling uses the most recent ERCOT day-ahead prices available for 2026, whereas the continental observed-price maps use January and July 2025, the latest common year available across the datasets. The Texas evaluation weeks are 1--7 January and 1--7 July 2026, representing winter and summer conditions, respectively. Dates are expressed in UTC, with prices for 8 January and 8 July used during the respective 24-hour execution extensions.

Data center locations and IT capacities are assembled from public facility information, including the Epoch AI Data Center Map~\cite{epochAI}. The modeled portfolios use configured capacity projections for 2030, together with site-specific reported or configured power usage effectiveness (PUE). These capacities describe the modeled computing infrastructure rather than verified operating demand at the named facilities.

Computing jobs are derived from the A10 subset of Alibaba's public spot-GPU cluster trace~\cite{duan2026GFS}, representing a relatively responsive workload regime. Different seven-day evaluation windows are assigned to the operator groups. Jobs overlapping each window, including previously submitted jobs that remain active, are retained subject to configured screening and sampling rules. Submission times are rounded down and execution durations rounded up to whole hours, with release and nominal completion times clipped to the original window. Resource requests equal requested CPUs multiplied by worker count and are treated as CPU-equivalent quantities. No new jobs arrive during the extension.

Computational capacity is calibrated from the retained trace's nominal resource-demand trajectory and configured utilization targets, excluding the first and last 24 hours where specified. Capacity is allocated among sites using configured shares and utilization targets, while individual job requests remain unchanged. These normalized capacities represent computing service rather than hardware inventories inferred from electrical ratings.

Transmission network inputs are obtained from PyPSA-USA~\cite{pypsausa}. Separate Texas and continental network representations provide the topology, generation, storage, background demand and renewable availability used in the analysis.

\subsection{Electricity-price magnitude and directional bias}

For comparison location $i$ and reference location $r$, the hourly electricity-price difference is: 
\begin{equation}
\Delta p_i^{(r)}(t)=p_i(t)-p_r(t).
\end{equation}
Positive values indicate that electricity is less expensive at the reference location. Positive and negative accumulated differences over the analyzed hours $\mathcal{T}$ are: 
\begin{align}
A_{+,i}^{(r)}
&=\sum_{t\in\mathcal{T}}\max\{\Delta p_i^{(r)}(t),0\}\Delta t,\\
A_{-,i}^{(r)}
&=\sum_{t\in\mathcal{T}}\max\{-\Delta p_i^{(r)}(t),0\}\Delta t,
\end{align}
where $\Delta t=1$~h.

Gross magnitude, directional bias and mean absolute price difference are:
\begin{align}
G_i^{(r)}&=A_{+,i}^{(r)}+A_{-,i}^{(r)},\\
D_i^{(r)}&=\frac{A_{+,i}^{(r)}-A_{-,i}^{(r)}}{G_i^{(r)}},\\
M_i^{(r)}&=\frac{G_i^{(r)}}{|\mathcal{T}|\Delta t}.
\end{align}
For prices expressed in \$/MWh, $G$ has units of \$/MW and $M$ has units of \$/MWh. Gross magnitude accumulates absolute price differences over the evaluation period; it is not the realized saving of a scheduled workload.

Directional bias approaches $+1$ when the accumulated advantage predominantly favors the reference location and $-1$ when it predominantly favors the comparison location. Values near zero indicate similar accumulated advantages in the two directions, provided that gross magnitude is non-zero. The measure weights the magnitude and duration of price differences, rather than counting how frequently their sign changes.

When $G=0$, directional bias is undefined. The implementation stores zero and separately flags the absence of a price difference; this value is not interpreted as evidence of balanced non-zero opportunities. 

\subsection{Job-level scheduling}

Each data center operator $o$ schedules a job set $\mathcal{J}_o$ across its site set $\mathcal{S}_o$. For job $j$, the processed release hour is $r_j$, the nominal completion hour is $d_j$, and the nominal resource request is $q_j$. Requested compute work is:
\begin{equation}
W_j=q_j(d_j-r_j),
\end{equation}
measured in vCPU-hour equivalents.

The availability indicator $a_{j,t}$ restricts execution to each job's permitted window, including any eligible extension, and prevents execution before release. The decision variables are integer resource allocations $x_{j,s,t}\geq0$ and continuous service rates $c_{j,s,t}\geq0$. They satisfy:
\begin{align}
x_{j,s,t}&\leq \overline{X}_j a_{j,t},\\
\underline{\kappa}_s x_{j,s,t}
&\leq c_{j,s,t}
\leq \overline{\kappa}_s x_{j,s,t},\\
\sum_{j\in\mathcal{J}_o}x_{j,s,t}
&\leq N_s^{\mathrm{cpu}},
\end{align}
where $\overline{X}_j$ is the processed per-site resource bound and $N_s^{\mathrm{cpu}}$ is site computational capacity. The service-rate bounds are $\underline{\kappa}_s=0.5$ and $\overline{\kappa}_s=1$.

Delivered work accumulates across sites and hours:
\begin{equation}
w_j=\sum_{s\in\mathcal{S}_o}\sum_{t\in\mathcal{H}}
c_{j,s,t}\Delta t.
\end{equation}
The price-responsive configuration permits execution to pause and resume, resource allocations to change, and work to be distributed among sites belonging to the same data center operator. The formulation represents divisible computing work and permits simultaneous allocations across sites.

Each job receives no more than its requested work, and each data center operator must deliver at least 95\% of its total requested work:
\begin{align}
0\leq w_j&\leq W_j,\\
\sum_{j\in\mathcal{J}_o}w_j
&\geq0.95\sum_{j\in\mathcal{J}_o}W_j.
\end{align}
This commitment is weighted by compute work, rather than by the number of completed jobs. The formulation permits different completion fractions among jobs.

When execution is restricted to one site per job, binary assignment variables $z_{j,s}$ impose:
\begin{align}
\sum_{s\in\mathcal{S}_o}z_{j,s}&=1,\\
x_{j,s,t}&\leq\overline{X}_j z_{j,s}.
\end{align}
These restrictions are removed when spatial reallocation is permitted. Requiring complete service instead replaces the minimum delivery requirement for each data center operator with $w_j=W_j$ for every job. 

The main price-responsive calculations do not assign explicit costs to migration, interruption or changes in site power. Communication bandwidth, data-transfer delays and application dependencies are not represented as additional constraints.

\subsection{Facility power, service revenue and electricity expenditure}

The service workload at site $s$ is:
\begin{equation}
L_{s,t}=\sum_j c_{j,s,t},
\qquad
\overline{L}_s=\overline{\kappa}_sN_s^{\mathrm{cpu}}.
\end{equation}
Facility electricity demand follows a linear utilization model:
\begin{equation}
P_{s,t}=\mathrm{PUE}_s
\left[
P_s^{\mathrm{idle}}+
\left(P_s^{\mathrm{busy}}-P_s^{\mathrm{idle}}\right)
\frac{L_{s,t}}{\overline{L}_s}
\right].
\end{equation}
Here $P_s^{\mathrm{busy}}$ is configured IT power at full utilization, and idle IT power is set to $P_s^{\mathrm{idle}}=0.2P_s^{\mathrm{busy}}$. PUE is applied once to convert IT power into facility power.

Data center operators maximize configured computing service revenue minus electricity expenditure:
\begin{equation}
\max_{x,c}
\left\{
\sum_{j,s,t}\pi_{j,s,t}c_{j,s,t}\Delta t
-C_o^{\mathrm{elec}}
\right\}.
\end{equation}
Retained jobs are assigned an illustrative service-value coefficient of 0.243~\$/vCPU-hour equivalent, informed by spot GPU pricing~\cite{google_spot_pricing} and scaled by the inverse sampling fraction when trace subsampling is applied.

Implemented electricity expenditure is:
\begin{align}
C_o^{\mathrm{elec}}
={}&
\sum_{s\in\mathcal{S}_o}\sum_{t\in\mathcal{H}}
p_{s,t}\mathrm{PUE}_s
\left(P_s^{\mathrm{busy}}-P_s^{\mathrm{idle}}\right)
\frac{L_{s,t}}{\overline{L}_s}\Delta t\\
&+
\sum_{s\in\mathcal{S}_o}\sum_{t\in\mathcal{T}_0}
p_{s,t}\mathrm{PUE}_sP_s^{\mathrm{idle}}\Delta t.
\end{align}

Because the objective includes delivered-service revenue and permits partial completion, expenditure differences reflect changes in delivered work as well as execution timing, location and resource allocation.

\subsection{Price-independent scheduling reference}

The price-independent reference uses first-come, first-served scheduling. For each data center operator, jobs are ordered by release hour, with job identifiers resolving ties. The first waiting job is assigned to a site that can accommodate its nominal resource request. Among eligible sites, the scheduler selects the site with the lowest computational utilization after assigning the job, with ties resolved by site identifier. If no site can accommodate the first waiting job, it remains at the head of the queue; later jobs are not used to backfill the available capacity. Once started, a job retains its site and nominal resource allocation until its requested work is delivered or its original execution window ends. This scheduling rule does not use electricity prices, permit migration or extend execution beyond the original window.

The same conversion from computing resource use to facility power is applied under price-independent scheduling. Workload inputs and site capacities are held fixed between the summer and winter price comparisons, leaving the reference demand profile unchanged. In the reported runs, price-independent scheduling delivers all represented requested compute work, whereas price-responsive scheduling delivers 95\% for each data center operator.

\subsection{Transmission dispatch and demand accounting}

The Texas network contains 1,890 lines, with data center sites assigned to configured nearby 345-kV buses. Linear optimal power flow evaluates prescribed computing demand by minimizing operating expenditure subject to nodal active-power balance, generation and storage constraints, and transmission-flow limits. The formulation does not evaluate reactive-power requirements, contingency security or dynamic stability.

The supplied winter background-demand series is retained in both seasons; summer preparation substitutes July solar and onshore-wind availability. High-cost generators at AC buses represent electricity load shedding at 100,000~\$/MWh, distinct from unexecuted computing work. 

For Texas evaluations containing explicitly modeled data center demand, an estimated embedded component is removed from native demand before scheduled facility power is added. At bus $b$,
\begin{equation}
P_{b,t}^{\mathrm{net}}
=P_{b,t}^{\mathrm{base}}
-P_{b,t}^{\mathrm{embedded}}
+P_{b,t}^{\mathrm{dc}}.
\end{equation}
For a positive native-demand peak,
\begin{equation}
P_{b,t}^{\mathrm{embedded}}
=
\min\left\{
P_{b,t}^{\mathrm{base}},
P_b^{\mathrm{inst}}
\frac{P_{b,t}^{\mathrm{base}}}
{\max_{\tau}P_{b,\tau}^{\mathrm{base}}}
\right\},
\end{equation}
where $P_b^{\mathrm{inst}}$ is the configured estimate of embedded data center capacity. The embedded component is zero when the native-demand peak is zero.

Both scheduling scenarios use the same embedded-demand adjustment. The reference without explicitly modeled data centers retains native demand unchanged and therefore does not remove computing consumption already embedded in the background data.

\subsection{Demand and congestion metrics}

Combined computing demand and the maximum absolute single-site hourly change are:
\begin{align}
P_t^{\mathrm{combined}}&=\sum_sP_{s,t},\\
R_{\max}^{\mathrm{site}}
&=\max_{s,t\geq1}
\frac{|P_{s,t}-P_{s,t-1}|}{\Delta t},
\end{align}
with $\Delta t=1$~h. Combined peak demand is $\max_tP_t^{\mathrm{combined}}$. 

For transmission line $\ell$, the implemented loading statistic is:
\begin{equation}
u_{\ell,t}=\frac{|f_{\ell,t}|}{F_\ell^{\mathrm{nom}}},
\end{equation}
where $f_{\ell,t}$ is the modeled active-power flow at the recorded line terminal and $F_\ell^{\mathrm{nom}}$ is the nominal line rating used by the reporting routine. A line-hour is classified as congested when the absolute line loading reaches at least 99.9\% of its nominal rating:
\begin{equation}
I_{\ell,t}=
\mathbf{1}\{u_{\ell,t}\geq0.999\}.
\end{equation}

Hourly counts and total congested line-hours are:
\begin{align}
N_t^{\mathrm{cong}}&=\sum_{\ell}I_{\ell,t},\\
H^{\mathrm{cong}}&=\sum_tN_t^{\mathrm{cong}}\Delta t.
\end{align}
The same line ratings and congestion threshold are used across the Texas scheduling comparisons.

Redistribution is evaluated by comparing the line-hour indicators under price-responsive scheduling, $I_{\ell,t}^{\mathrm{PR}}$, with those under price-independent scheduling, $I_{\ell,t}^{\mathrm{PI}}$. Newly congested and relieved line-hours are:
\begin{align}
H^{\mathrm{new}}
&=\sum_{\ell,t}
I_{\ell,t}^{\mathrm{PR}}
\left(1-I_{\ell,t}^{\mathrm{PI}}\right)\Delta t,\\
H^{\mathrm{relieved}}
&=\sum_{\ell,t}
I_{\ell,t}^{\mathrm{PI}}
\left(1-I_{\ell,t}^{\mathrm{PR}}\right)\Delta t.
\end{align}
Consequently,
\begin{equation}
H_{\mathrm{PR}}^{\mathrm{cong}}
-H_{\mathrm{PI}}^{\mathrm{cong}}
=H^{\mathrm{new}}-H^{\mathrm{relieved}}.
\end{equation}

\subsection{Continental marginal prices and price feedback}
\label{subsec:methods_feedbacks}

In the continental network, data center sites are assigned to their geographically nearest load buses, with facility demand summed at shared buses. Computing-demand scenarios add these profiles to native demand, while the reference without data centers retains native demand alone.

For each prescribed demand profile, linear optimal power flow returns marginal electricity prices from the dual variables of the nodal active-power balance constraints. These prices are conditional marginal supply costs within the represented dispatch model, rather than reconstructed electricity-market settlements.

Price-difference magnitude and directional bias are calculated using the same definitions as for the observed-price analysis. Mean gross magnitude is averaged across network buses relative to the Texas bus hosting the largest number of modeled data center sites.

The mean absolute difference between data center site prices is calculated over all unordered pairs of the 48 sites and all evaluated hours:
\begin{equation}
M_{\mathrm{sites}}
=
\frac{1}{|\mathcal{T}_0|\binom{48}{2}}
\sum_{t\in\mathcal{T}_0}
\sum_{s<s'}
|p_{b(s),t}-p_{b(s'),t}|,
\end{equation}
where $b(s)$ is the network bus assigned to site $s$. Sites sharing a bus are retained as distinct sites, so their pairwise price difference is zero. Directional-pattern similarity is summarized by Pearson correlation between the aligned bus-level directional-bias arrays, using the stated zero-spread storage convention. 

Computing schedules and network dispatch are updated iteratively. At each iteration, operators reschedule workloads using the preceding dispatch prices, and the resulting facility demand profiles enter a new network optimization, with background demand and other inputs held fixed. The final hourly dispatch price is repeated over the execution extension. Summer spatial price maps and battery arbitrage calculations use prices obtained after the sixth update.

\subsection{Battery energy arbitrage}
\label{subsec:methods_battery}

A price-taking battery is optimized separately at each network bus using the summer prices under price-independent and updated price-responsive computing demand. Each battery has a power capacity of $\overline{P}^{\mathrm{bat}}=1$~MW, an energy capacity of $\overline{E}^{\mathrm{bat}}=4$~MWh, and efficiencies $\eta_{\mathrm{ch}}=\eta_{\mathrm{dis}}=\sqrt{0.90}$.

For each bus $b$, charging power $P_{b,t}^{\mathrm{ch}}$, discharging power $P_{b,t}^{\mathrm{dis}}$ and stored energy $E_{b,t}$ are optimized over $T=168$ hours. Potential arbitrage value is:
\begin{equation}
V_b=
\max
\sum_{t=0}^{T-1}
p_{b,t}
\left(
P_{b,t}^{\mathrm{dis}}-P_{b,t}^{\mathrm{ch}}
\right)\Delta t,
\end{equation}
where $p_{b,t}$ is the prescribed marginal electricity price and $\Delta t=1$~h. Stored energy evolves according to:
\begin{equation}
E_{b,t+1}
=
E_{b,t}
+\eta_{\mathrm{ch}}P_{b,t}^{\mathrm{ch}}\Delta t
-\frac{P_{b,t}^{\mathrm{dis}}\Delta t}{\eta_{\mathrm{dis}}},
\end{equation}
subject to:
\begin{align}
0&\leq E_{b,t}\leq\overline{E}^{\mathrm{bat}},\\
E_{b,0}&=E_{b,T}=0.5\overline{E}^{\mathrm{bat}}.
\end{align}
A binary variable $u_{b,t}$ prevents simultaneous charging and discharging:
\begin{align}
0&\leq P_{b,t}^{\mathrm{ch}}
\leq\overline{P}^{\mathrm{bat}}u_{b,t},\\
0&\leq P_{b,t}^{\mathrm{dis}}
\leq\overline{P}^{\mathrm{bat}}(1-u_{b,t}),\\
u_{b,t}&\in\{0,1\}.
\end{align}

The complete weekly price series is known to the battery optimization. The objective measures electricity sales revenue minus charging expenditure and excludes installation, maintenance and battery degradation costs. All battery specifications and operating constraints are held fixed between the two price inputs.

The change in potential arbitrage value is:
\begin{equation}
\Delta v_b
=
\frac{
V_b^{\mathrm{PR}}-V_b^{\mathrm{PI}}
}{
\overline{P}^{\mathrm{bat}}
},
\end{equation}
where PR and PI denote prices under price-responsive and price-independent computing demand, respectively. Values are reported per MW of battery power capacity over the evaluated week. Negative values indicate reduced arbitrage opportunities under price-responsive computing demand.

An illustrative capital recovery calculation uses observed day-ahead prices from January and July 2026. Baseline annualized arbitrage revenue is calculated as the mean optimized revenue across complete 168-hour blocks multiplied by 52. This revenue is scaled by the ratio of mean daily price ranges under price-responsive versus price-independent computing demand in the modeled summer week, where each daily range is the maximum minus the minimum hourly price. The ratios are 0.694, 0.927 and 1.000 at the selected Texas, Michigan and California nodes, respectively. This proportional adjustment is illustrative rather than the result of battery optimization against an adjusted annual price series. Installation cost is set to 326~\$/kWh, corresponding to 1.304 million dollars for the modeled 1~MW/4~MWh battery, according to the NREL cost projection~\cite{cole2025cost}. Cumulative revenue assumes constant annual earnings, and the simple capital recovery period is installation cost divided by annual revenue, with a 15-year operating life shown for reference. Discounting, maintenance, degradation and replacement costs are excluded.

\subsection{Aggregate scheduling models}
\label{subsec:methods_agg_model}

The aggregate models replace individual jobs with continuous site-power decisions, omitting release times, execution windows and job-level resource allocations. They use the same site-associated prices as the job-level comparison.

Let $P_{s,t}^{\mathrm{job}}$ denote the reference facility power profile from job-level scheduling. Each aggregate model constrains hourly power at each site to remain between the minimum and maximum values of its reference profile over the original 168 hours. These bounds describe the observed facility power range rather than independently estimated limits on feasible computing demand.

The site-energy model, labeled AGG Site, preserves each site's reference electricity consumption:
\begin{equation}
\sum_{t\in\mathcal{T}_0}P_{s,t}\Delta t
=
\sum_{t\in\mathcal{T}_0}P_{s,t}^{\mathrm{job}}\Delta t
\quad\text{for each }s.
\end{equation}
The model labeled AGG Operator preserves total electricity consumption across the sites belonging to each data center operator:
\begin{equation}
\sum_{s\in\mathcal{S}_o}\sum_{t\in\mathcal{T}_0}
P_{s,t}\Delta t
=
\sum_{s\in\mathcal{S}_o}\sum_{t\in\mathcal{T}_0}
P_{s,t}^{\mathrm{job}}\Delta t
\quad\text{for each }o.
\end{equation}
It can therefore redistribute the matched energy across both sites and hours. AGG Continuous imposes only the power bounds, without an energy requirement; its lower expenditure may therefore reflect reduced consumption rather than equivalent service delivery.

The aggregate objective is: 
\begin{equation}
\min_P\sum_s\sum_{t\in\mathcal{T}_0}
p_{s,t}P_{s,t}\Delta t.
\end{equation}
No additional hourly ramp constraints are imposed. Energy matching applies to electricity consumption within $\mathcal{T}_0$, rather than computing service delivery over $\mathcal{H}$.

The summary figure reports arithmetic means of the summer and winter metrics. Percentage differences in expenditure, peak demand and maximum absolute hourly demand change are calculated relative to the corresponding job-level means. Demand changes are evaluated both at the single site of operator A and across all sites. 
Mean absolute error (MAE) measures hourly demand deviations from the job-level reference:
\begin{equation}
\mathrm{MAE}
=
\frac{1}{|\mathcal{T}_0|}
\sum_{t\in\mathcal{T}_0}
\left|P_t^{\mathrm{AGG}}-P_t^{\mathrm{job}}\right|.
\end{equation}
For site-level MAE, $P_t$ is demand at the site of operator A; for combined MAE, $P_t=\sum_sP_{s,t}$. Seasonal MAEs are averaged and reported in GW.

\subsection{Regional demand maps}

For each geographic reference location $i$, an average 24-hour demand profile $\overline{P}_i(h)$ is constructed by averaging the available demand observations at each UTC hour within the selected month. Its local daily peak is defined as ${P_i^\star}$. The fraction below the local peak is:
\begin{equation}
q_i(h)=
\frac{P_i^\star-\overline{P}_i(h)}{P_i^\star}.
\end{equation}
Smaller values indicate demand closer to that location's own peak. For visualization, fractions are clipped to $[0.001,0.2]$ and transformed using $\log_{10}$ before spatial interpolation. 

\subsection{Computation and reproducibility}

Calculations are performed on a workstation equipped with an Intel Core i9-14900K processor and 64~GB RAM. Job-level scheduling models are implemented in Pyomo and solved with Gurobi. Battery arbitrage models use the Gurobi Python interface, and network dispatch is implemented in PyPSA. Statistics are calculated from numerical outputs; spatial interpolation is used only for visualization.
\section*{Data availability}
The processed demand profiles, price summaries, site metadata, scheduled power profiles and network outputs underlying the figures are available in the 
\href{https://github.com/Shijie-Pan/The-Cheap-Hours-Dont-Stay-Cheap-Price-Chasing-Data-Centers-Reshape-Grids-and-Markets}{GitHub repository} and archived on Zenodo at \url{https://doi.org/10.5281/zenodo.23507446}. 
The Alibaba spot-GPU trace and day-ahead electricity prices are available from their respective providers and are not redistributed~\cite{alibaba_cluster_trace,gridstatus_data}.

\section*{Code availability}
The scheduling models, network inputs and figure scripts are available in the 
\href{https://github.com/Shijie-Pan/The-Cheap-Hours-Dont-Stay-Cheap-Price-Chasing-Data-Centers-Reshape-Grids-and-Markets}{GitHub repository}. The version accompanying this paper is archived on Zenodo at \url{https://doi.org/10.5281/zenodo.23507446}. 
Instructions for reproducing the figures and rerunning the calculations are provided in the repository README. Rerunning the optimization requires a local Gurobi license and the third-party workload and price data described there.

\section*{Acknowledgements}
This research was supported by King Abdullah University of Science and Technology (KAUST) through the KAUST Research Funding and Services (RFS) under Award No.~ORFS-KAUST-RDIA-2023-5720 and through the Cisco--KAUST AI Institute under the project ``PowerAware AI: Energy-Aware Workload Management for Cost-Efficient AI Data Center Operations'' (RFS Award No.~7542). C.K.\ acknowledges the Donald D.\ Harrington Faculty Fellows Program for partial support of this research.

\section*{Author contributions}
S.P.\ designed the study, performed the experiments and drafted the manuscript. Z.A.A.\ and C.K.\ supervised the research and contributed to interpretation and manuscript revision. All authors approved the final manuscript.

\section*{Competing interests}
The authors declare no competing interests.

\section*{Additional information}
\textbf{Supplementary information} is available for this paper.
\textbf{Correspondence} and requests for materials should be addressed to Shijie Pan.

\bibliography{refs}

@misc{ny_eo62,
  author       = {{Office of the Governor of New York}},
  title        = {Executive Order No. 62: Establishing a Temporary Moratorium on Data Centers in {New York} While the State Develops Higher Standards for Data Center Development and Benefits Blueprint to Support Localities},
  year         = {2026},
  month        = jul,
  day          = {14},
  howpublished = {\url{https://www.governor.ny.gov/executive-order/no-62-establishing-temporary-moratorium-data-centers-new-york-while-state-develops}},
  note         = {Accessed: Oct. 20, 2026}
}

@online{bundesnetzagentur2026agnes,
  author  = {{Bundesnetzagentur}},
  title   = {Bundesnetzagentur Presents Latest Considerations on Reform of Electricity Network Tariff System},
  year    = {2026},
  month   = may,
  day     = {27},
  url     = {https://www.bundesnetzagentur.de/SharedDocs/Pressemitteilungen/EN/2026/20260527_agnes.html},
  urldate = {2026-10-18},
  note    = {Accessed: Oct. 20, 2026}
}

@misc{epochAI,
  author  = {{Epoch AI}},
  title   = {{AI Data Center Map}},
  url     = {https://epoch.ai/data/ai-data-centers/map},
  urldate = {2026-10-18},
  note    = {Accessed: Oct. 20, 2026}
}

@techreport{shehabi20242024,
  author = {Shehabi, Arman and Smith, Sarah Josephine and Hubbard, Alex and Newkirk, Alexander and Lei, Nuoa and Siddik, Md AbuBakar and Holecek, Billie and Koomey, Jonathan G. and Masanet, Eric R. and Sartor, Dale A.},
  title = {{2024 United States Data Center Energy Usage Report}},
  institution = {Lawrence Berkeley National Laboratory},
  year = {2024},
  doi = {10.71468/P1WC7Q},
  url = {https://eta-publications.lbl.gov/publications/2024-lbnl-data-center-energy-usage-report}
}

@article{crozier2025potential,
  author = {Crozier, Constance and Liska, Matthew},
  title = {The potential of data center energy demand to provide grid flexibility},
  journal = {Current Sustainable/Renewable Energy Reports},
  year = {2025},
  doi = {10.1007/s40518-025-00258-9},
  volume = {12},
  number = {1},
  pages = {12}
}

@article{colangelo2026ai,
  title={{AI data centres as grid-interactive assets}},
  author={Colangelo, Philip and Coskun, Ayse K and Megrue, Jack and Roberts, Ciaran and Sengupta, Shayan and Sivaram, Varun and Tiao, Ethan and Vijaykar, Aroon and Williams, Chris and Wilson, Daniel C and others},
  journal={Nature Energy},
  volume={11},
  number={2},
  pages={254--261},
  year={2026},
  publisher={Nature Publishing Group UK London}
}

@article{zhou2024energy,
  author = {Zhou, Shibo and Zhou, Ming and Wu, Zhaoyuan and Wang, Yuyang and Li, Gengyin},
  title = {{Energy-aware coordinated operation strategy of geographically distributed data centers}},
  journal = {International Journal of Electrical Power \& Energy Systems},
  year = {2024},
  doi = {10.1016/j.ijepes.2024.110032},
  volume = {159},
  pages = {110032}
}

@article{cao2024managing,
  title={{Managing data center cluster as non-wire alternative: A case in balancing market}},
  author={Cao, Yujie and Cao, Fang and Wang, Yajing and Wang, Jianxiao and Wu, Lei and Ding, Zhaohao},
  journal={Applied Energy},
  volume={360},
  pages={122769},
  year={2024},
  publisher={Elsevier}
}

@misc{google_spot_pricing,
  author       = {{Google Cloud}},
  title        = {{Spot {VM}s Pricing}},
  year         = {2026},
  howpublished = {\url{https://cloud.google.com/spot-vms/pricing}}
}

@article{kim2026flexibility,
  author = {Kim, Dongjoo and Dong, Lin and Xie, Le},
  title = {{Flexibility-aware framework for efficient planner-initiated siting of data center}},
  journal = {Nature Communications},
  volume = {17},
  pages = {6512},
  year = {2026},
  doi = {10.1038/s41467-026-72324-9}
}

@misc{alibaba_cluster_trace,
 author={{Alibaba}}, 
 title={{Traces for {AI} jobs leveraging spot {GPU} resources}},
 howpublished={\url{https://github.com/alibaba/clusterdata/tree/master/cluster-trace-v2026-spot-gpu}}, 
 year={2026}
}

@misc{ERCOTLargeLoad,
  author       = {{Electric Reliability Council of Texas}},
  title        = {{Planning Guide: Section 9, Large Load Additions at New or Modification of Existing Load Interconnections}},
  year         = {2026},
  howpublished={\url{https://www.ercot.com/files/docs/2026/03/30/April-1-2026-Planning-Guide.pdf}}
}

@inproceedings{duan2026gfs,
author = {Duan, Jiaang and Xu, Shenglin and Qian, Shiyou and Yang, Dingyu and Wang, Kangjin and Liao, Chenzhi and Yu, Yinghao and Hua, Qin and Hu, Hanwen and Wang, Qi and Wu, Wenchao and Bao, Dongqing and Lu, Tianyu and Cao, Jian and Xue, Guangtao and Yang, Guodong and Zhang, Liping and Chen, Gang},
title = {{GFS}: A Preemption-aware Scheduling Framework for {GPU} Clusters with Predictive Spot Instance Management},
year = {2025},
isbn = {9798400721656},
publisher = {Association for Computing Machinery},
address = {New York, NY, USA},
url = {https://doi.org/10.1145/3760250.3762231},
doi = {10.1145/3760250.3762231},
editor = {Tang, Xulong and Narayanan, Vijaykrishnan and Lee, Benjamin C. and Xu, Harry and Silberstein, Mark},
booktitle = {Proceedings of the 31st ACM International Conference on Architectural Support for Programming Languages and Operating Systems, Volume 1},
pages = {117--131},
numpages = {15},
location = {Pittsburgh, PA, USA},
series = {ASPLOS '26}
}

@inproceedings{pavlenko2024vertically,
author = {Pavlenko, Anna and Cahoon, Joyce and Zhu, Yiwen and Kroth, Brian and Nelson, Michael and Carter, Andrew and Liao, David and Wright, Travis and Camacho-Rodr{\'i}guez, Jes{\'u}s and Saur, Karla},
title = {Vertically Autoscaling Monolithic Applications with {CaaSPER}: Scalable Container-as-a-Service Performance Enhanced Resizing Algorithm for the Cloud},
year = {2024},
isbn = {9798400704222},
publisher = {Association for Computing Machinery},
address = {New York, NY, USA},
url = {https://doi.org/10.1145/3626246.3653378},
doi = {10.1145/3626246.3653378},
editor = {Barcel{\'o}, Pablo and S{\'a}nchez-Pi, Nayat and Meliou, Alexandra and Sudarshan, S.},
booktitle = {Companion of the 2024 International Conference on Management of Data},
pages = {241--254},
numpages = {14},
location = {Santiago AA, Chile},
series = {SIGMOD '24},
}

@article{dayarathna2015data,
  author={Dayarathna, Miyuru and Wen, Yonggang and Fan, Rui},
  journal={IEEE Communications Surveys \& Tutorials}, 
  title={Data center energy consumption modeling: A survey},
  year={2016},
  volume={18},
  number={1},
  pages={732--794},

  doi={10.1109/COMST.2015.2481183}}

@misc{pypsausa,
  author = {Tehranchi, Kamran and Barnes, Trevor and Frysztacki, Martha and Azevedo, Ines},
  title = {{PyPSA-USA: A Flexible Open-Source Energy System Model and Optimization Tool for the United States}},
  year = {2025},
  doi = {10.2139/ssrn.5029120},
  url = {https://doi.org/10.2139/ssrn.5029120}
}

@techreport{iea2025energyai,
  author = {{International Energy Agency}},
  title = {{Energy and AI}},
  institution = {International Energy Agency},
  year = {2025},
  url = {https://www.iea.org/reports/energy-and-ai}
}

@techreport{iea2026questions,
  author = {{International Energy Agency}},
  title = {{Key Questions on Energy and AI}},
  institution = {International Energy Agency},
  year = {2026},
  url = {https://www.iea.org/reports/key-questions-on-energy-and-ai}
}

@article{radovanovic2022carbon,
  author={Radovanovi{\'c}, Ana and Koningstein, Ross and Schneider, Ian and Chen, Bokan and Duarte, Alexandre and Roy, Binz and Xiao, Diyue and Haridasan, Maya and Hung, Patrick and Care, Nick and Talukdar, Saurav and Mullen, Eric and Smith, Kendal and Cottman, MariEllen and Cirne, Walfredo},
  journal={IEEE Transactions on Power Systems}, 
  title={Carbon-aware computing for datacenters},
  year={2023},
  volume={38},
  number={2},
  pages={1270--1280},

  doi={10.1109/TPWRS.2022.3173250}}

@article{nature2026planning,
  title = {{Rethink data centre planning}},
  journal = {Nature Energy},
  year = {2026},
  doi = {10.1038/s41560-026-02146-y},
  key = {Nature Energy},
  note = {Editorial},
  volume = {11},
  pages = {1383}
}

@article{cao2022assessment,
  author = {Cao, Yujie and Cheng, Ming and Zhang, Sufang and Mao, Hongju and Wang, Peng and Li, Chao and Feng, Yihui and Ding, Zhaohao},
  title = {{Data-driven flexibility assessment for internet data center towards periodic batch workloads}},
  journal = {Applied Energy},
  year = {2022},
  doi = {10.1016/j.apenergy.2022.119665},
  volume = {324},
  pages = {119665}
}

@misc{zhou2024hpc,
  author = {Zhou, Yihong and Paredes, Angel and Essayeh, Chaimaa and Morstyn, Thomas},
  title = {{AI}-focused {HPC} data centers can provide more power grid flexibility and at lower cost},
  year = {2024},
  howpublished = {Preprint at \url{https://arxiv.org/abs/2410.17435}},
  doi = {10.48550/arXiv.2410.17435}
}

@techreport{cole2025cost,
  author       = {Cole, Wesley and Ramasamy, Vignesh and Turan, Merve},
  title        = {Cost Projections for Utility-Scale Battery Storage: 2025 Update},
  institution  = {National Renewable Energy Laboratory (NREL), Golden, CO (United States)},
  doi          = {10.2172/2583471},
  url          = {https://www.osti.gov/biblio/2583471},
  place        = {United States},
  year         = {2025},
  month        = {06}}

@article{zhang2022exploring,
  author={Zhang, Weiqi and Roald, Line and Zavala, Victor},
  journal={IEEE Transactions on Sustainable Computing}, 
  title={Exploring the impacts of power grid signals on data center operations using a receding-horizon scheduling model},
  year={2023},
  volume={8},
  number={2},
  pages={245--256},

  doi={10.1109/TSUSC.2022.3224668}}

@article{zhang2020flexibility,
  title={{Flexibility from networks of data centers: A market clearing formulation with virtual links}},
  author={Zhang, Weiqi and Roald, Line A and Chien, Andrew A and Birge, John R and Zavala, Victor M},
  journal={Electric Power Systems Research},
  volume={189},
  pages={106723},
  year={2020},
  publisher={Elsevier}
}

@article{zhang2022remunerating,
  title={{Remunerating space--time, load-shifting flexibility from data centers in electricity markets}},
  author={Zhang, Weiqi and Zavala, Victor M},
  journal={Applied Energy},
  volume={326},
  pages={119930},
  year={2022},
  publisher={Elsevier}
}

@article{chen2025spatial,
  author = {Chen, Boyu and Che, Yanbo and Takci, Mehmet and Qadrdan, Meysam and Zhou, Yue},
  title = {{Spatial flexibility provision from geographically dispersed data centers enabling coordinated operation of multi-local flexibility markets}},
  journal = {IEEE Transactions on Smart Grid},
  volume = {17},
  number = {2},
  pages = {1371--1381},
  year = {2026},
  doi = {10.1109/TSG.2025.3648617}
}

@article{li2023computation,
  author={Li, Chuyi and Zheng, Kedi and Guo, Hongye and Kang, Chongqing and Chen, Qixin},
  journal={IEEE Transactions on Smart Grid}, 
  title={Computation-power coupled modeling for {IDCs} and collaborative optimization in {ADNs}},
  year={2024},
  volume={15},
  number={3},
  pages={2762--2775},

  doi={10.1109/TSG.2023.3321376}}

@article{zhang2021hpc,
  author={Zhang, Yijia and Wilson, Daniel Curtis and Paschalidis, Ioannis Ch. and Coskun, Ayse K.},
  journal={IEEE Transactions on Sustainable Computing}, 
  title={{HPC} data center participation in demand response: An adaptive policy with {QoS} assurance},
  year={2022},
  volume={7},
  number={1},
  pages={157--171},

  doi={10.1109/TSUSC.2021.3077254}}

@misc{pudl_data,
  author = {{Catalyst Cooperative}},
  title = {{Public Utility Data Liberation (PUDL): Data Viewer}},
  url = {https://data.catalyst.coop/},
  urldate = {2026-10-18},
  note    = {Accessed: Oct. 20, 2026}
}

@misc{ferc714,
  author = {{Federal Energy Regulatory Commission}},
  title = {{Form No. 714---Annual Electric Balancing Authority Area and Planning Area Report}},
  url = {https://www.ferc.gov/industries-data/electric/general-information/electric-industry-forms/form-no-714-annual-electric/overview},
  urldate = {2026-10-18},
  note    = {Accessed: Oct. 20, 2026}
}

@misc{eia930,
  author = {{U.S. Energy Information Administration}},
  title = {{EIA-930: Hourly and Daily Balancing Authority Operations Report}},
  url = {https://www.eia.gov/survey/#eia-930},
  urldate = {2026-10-18},
  note    = {Accessed: Oct. 20, 2026}
}

@misc{gridstatus_data,
  author = {{GridStatus}},
  title = {{Grid Status}},
  url = {https://www.gridstatus.io/live},
  urldate = {2026-10-18},
  note    = {Accessed: Oct. 20, 2026}
}

@article{riepin2025spatio,
  title={{Spatio-temporal load shifting for truly clean computing}},
  author={Riepin, Iegor and Brown, Tom and Zavala, Victor M},
  journal={Advances in Applied Energy},
  volume={17},
  pages={100202},
  year={2025},
  publisher={Elsevier}
}

@article{zhang2025mitigating,
  title={{Mitigating power grid impact from proactive data center workload shifts: A coordinated scheduling strategy integrating synergistic traffic-data-power networks}},
  author={Zhang, Yuanshi and Zou, Bokang and Jin, Xu and Luo, Yifu and Song, Meng and Ye, Yujian and Hu, Qinran and Chen, Qirui and Zambroni, Antonio Carlos},
  journal={Applied Energy},
  volume={377},
  pages={124697},
  year={2025},
  publisher={Elsevier}
}

@article{sioshansi2009estimating,
  author  = {Sioshansi, Ramteen and Denholm, Paul and Jenkin, Thomas and Weiss, Jurgen},
  title   = {Estimating the value of electricity storage in {PJM}: Arbitrage and some welfare effects},
  journal = {Energy Economics},
  volume  = {31},
  number  = {2},
  pages   = {269--277},
  year    = {2009}
}

@article{lamp2022large,
  author  = {Lamp, Stefan and Samano, Mario},
  title   = {Large-scale battery storage, short-term market outcomes, and arbitrage},
  journal = {Energy Economics},
  volume  = {107},
  pages   = {105786},
  year    = {2022}
}

\clearpage
\clearpage
\setcounter{figure}{0}
\renewcommand{\figurename}{Supplementary Fig.}

\begin{figure*}[h]
\centering
\includegraphics[width=\textwidth]{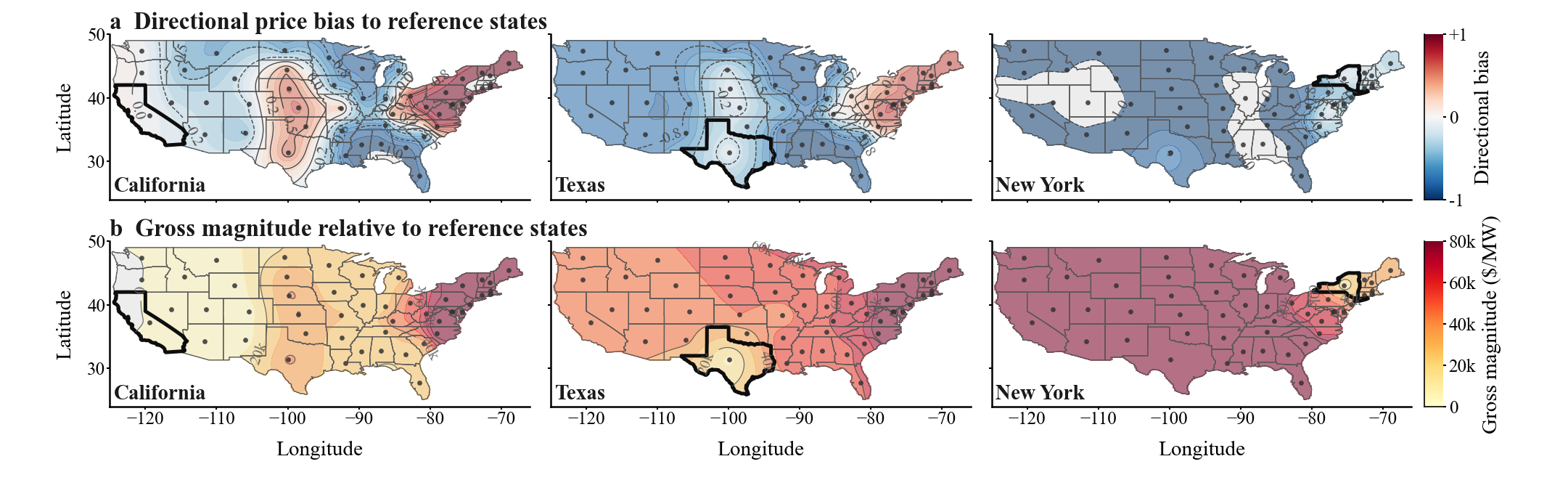}
\caption{\textbf{Winter magnitude and directional bias of regional day-ahead price differences.}
January 2025 results corresponding to Fig.~2.}
\label{fig:supp_winter_DG}
\end{figure*}

\begin{figure*}[h]
\centering
\includegraphics[width=\textwidth]{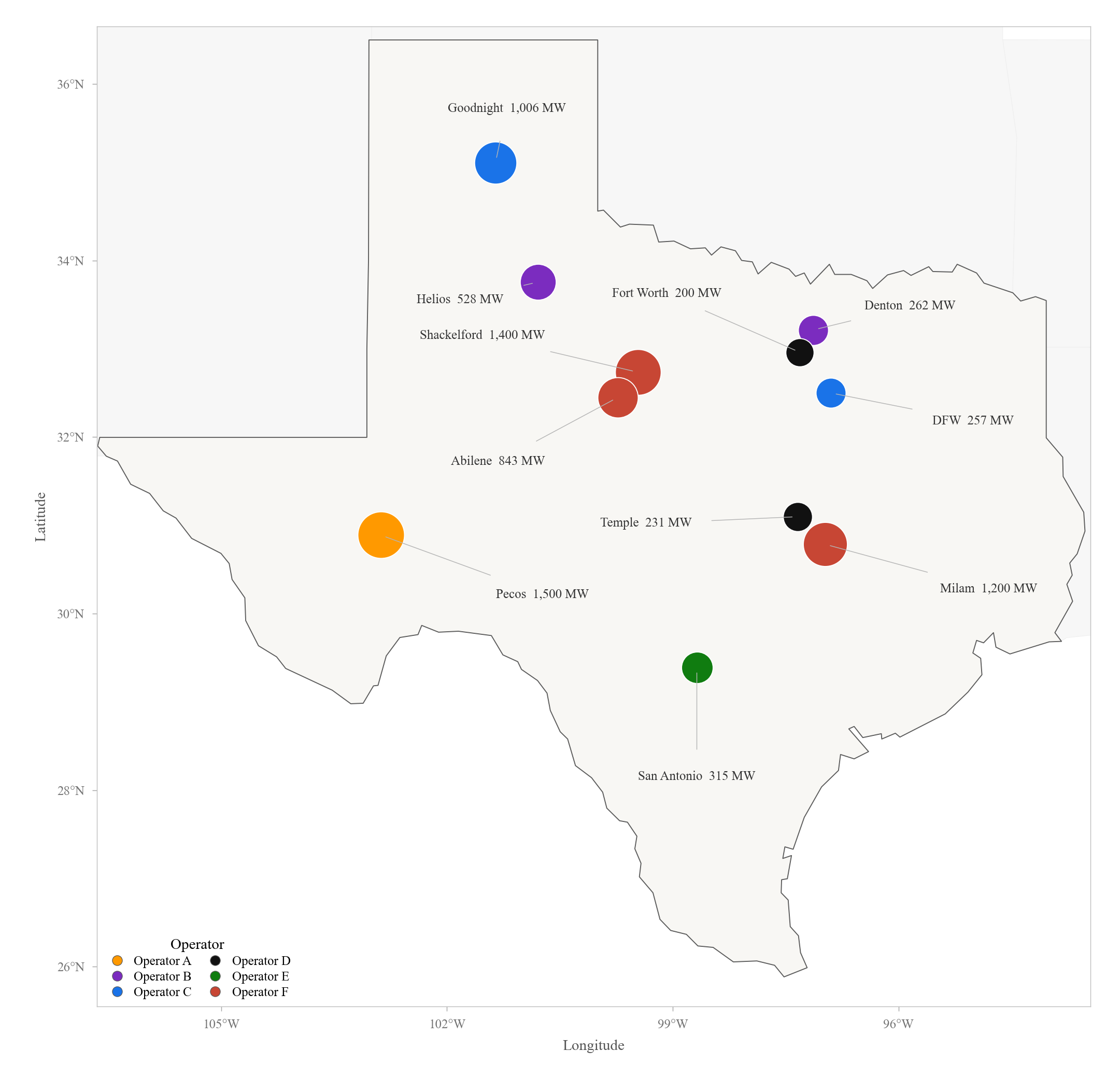}
\caption{\textbf{Locations and modeled capacities of the 11 Texas data center sites used in the scheduling experiments.}}
\label{fig:supp-campus-map}
\end{figure*}

\begin{figure*}[h]
\centering
\includegraphics[width=\textwidth]{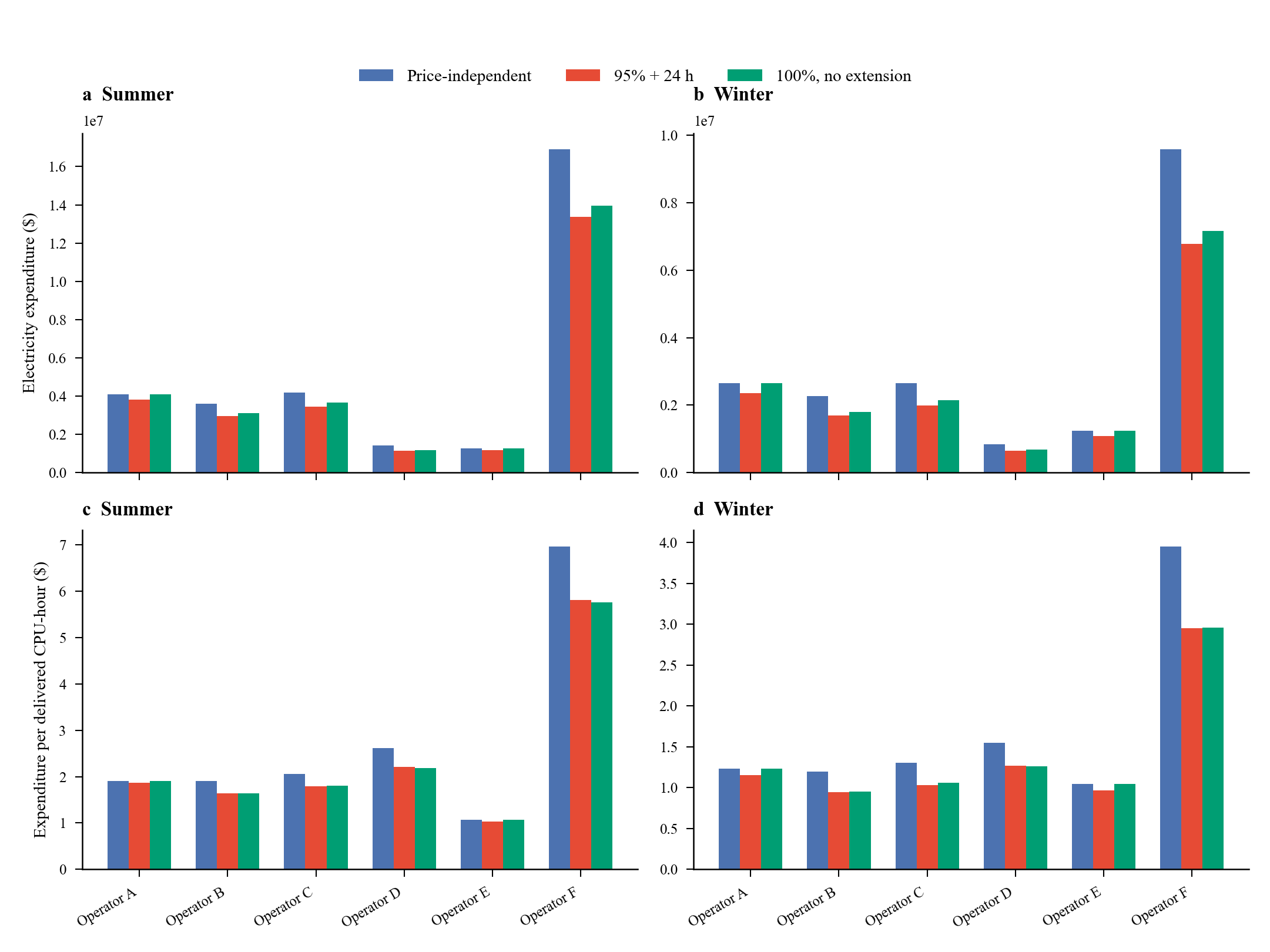}
\caption{\textbf{Electricity savings under different service requirements.}
Total electricity expenditure (top) and expenditure per delivered vCPU-hour (bottom) for the six modeled data center operators under summer (left) and winter (right) prices. Price-independent scheduling is compared with price-responsive scheduling requiring 95\% delivery with a 24-hour extension or full delivery within the original job windows.}
\label{fig:supp-service-conditions}
\end{figure*}

\begin{figure*}[h]
\centering
\includegraphics[width=\textwidth]{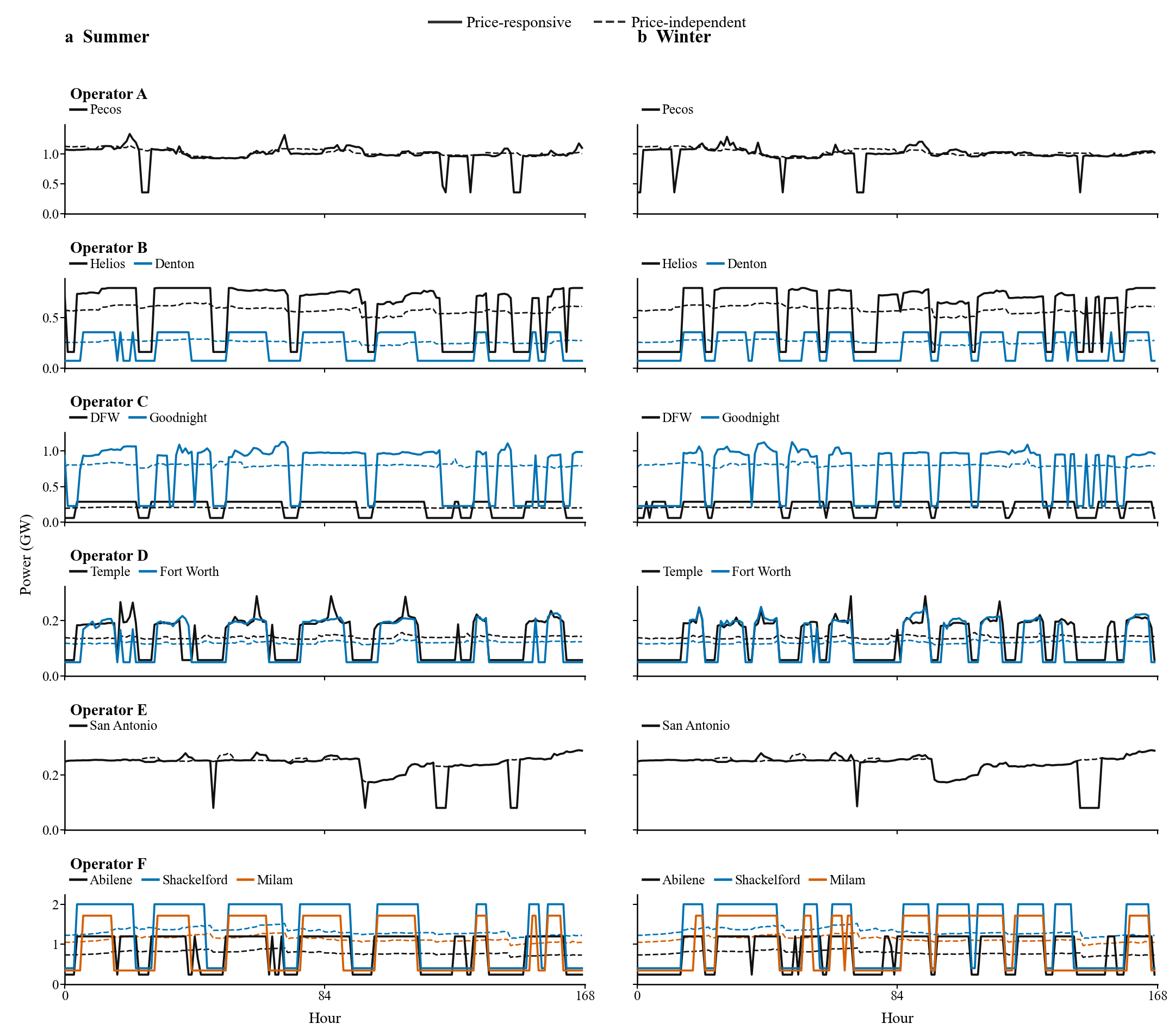}
\caption{\textbf{Site-level electricity demand under price-independent and price-responsive scheduling.}
Hourly demand at the 11 modeled Texas data center sites under summer and winter price conditions.}
\label{fig:supp-site-trace}
\end{figure*}

\begin{figure*}[h]
\centering
\includegraphics[width=\textwidth]{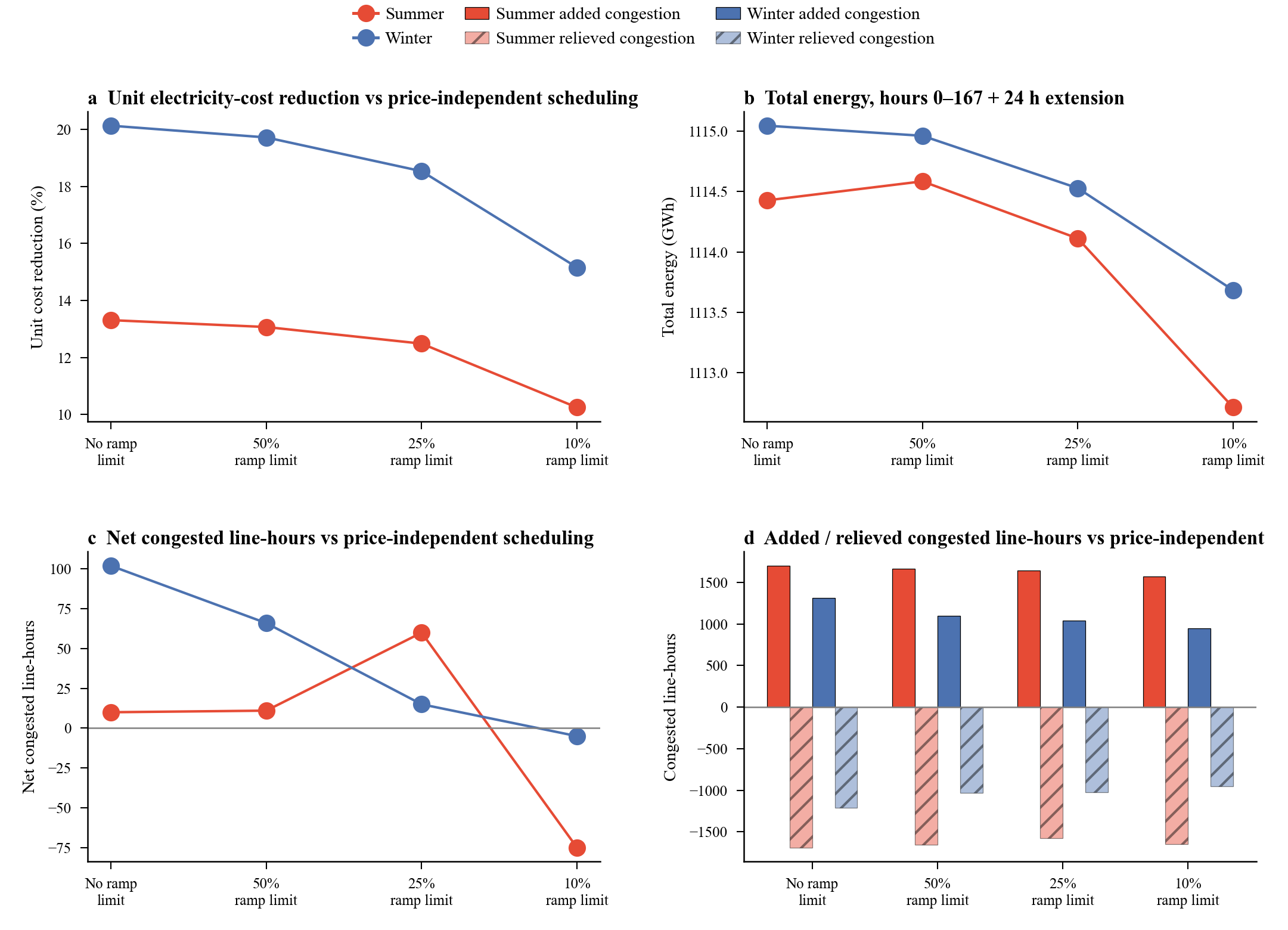}
\caption{\textbf{Electricity savings and transmission congestion under site-level ramp limits.}
All price-responsive cases require at least 95\% work delivery and allow a 24-hour extension. 
Ramp limits constrain hourly power changes to 50\%, 25\% or 10\% of each site's modeled facility power upper bound.
\textbf{a}, Reduction in electricity expenditure per delivered vCPU-hour relative to price-independent scheduling.
\textbf{b}, Total electricity consumption over the 192-hour scheduling horizon.
\textbf{c}, Net change in congested line-hours relative to price-independent scheduling.
\textbf{d}, Newly congested and relieved line-hours, shown as positive and negative values, respectively.
All ramp-limit cases are compared with the price-independent reference.}
\label{fig:supp-ramp-sensitivity}
\end{figure*}

\begin{figure*}[h]
\centering
\includegraphics[width=\textwidth]{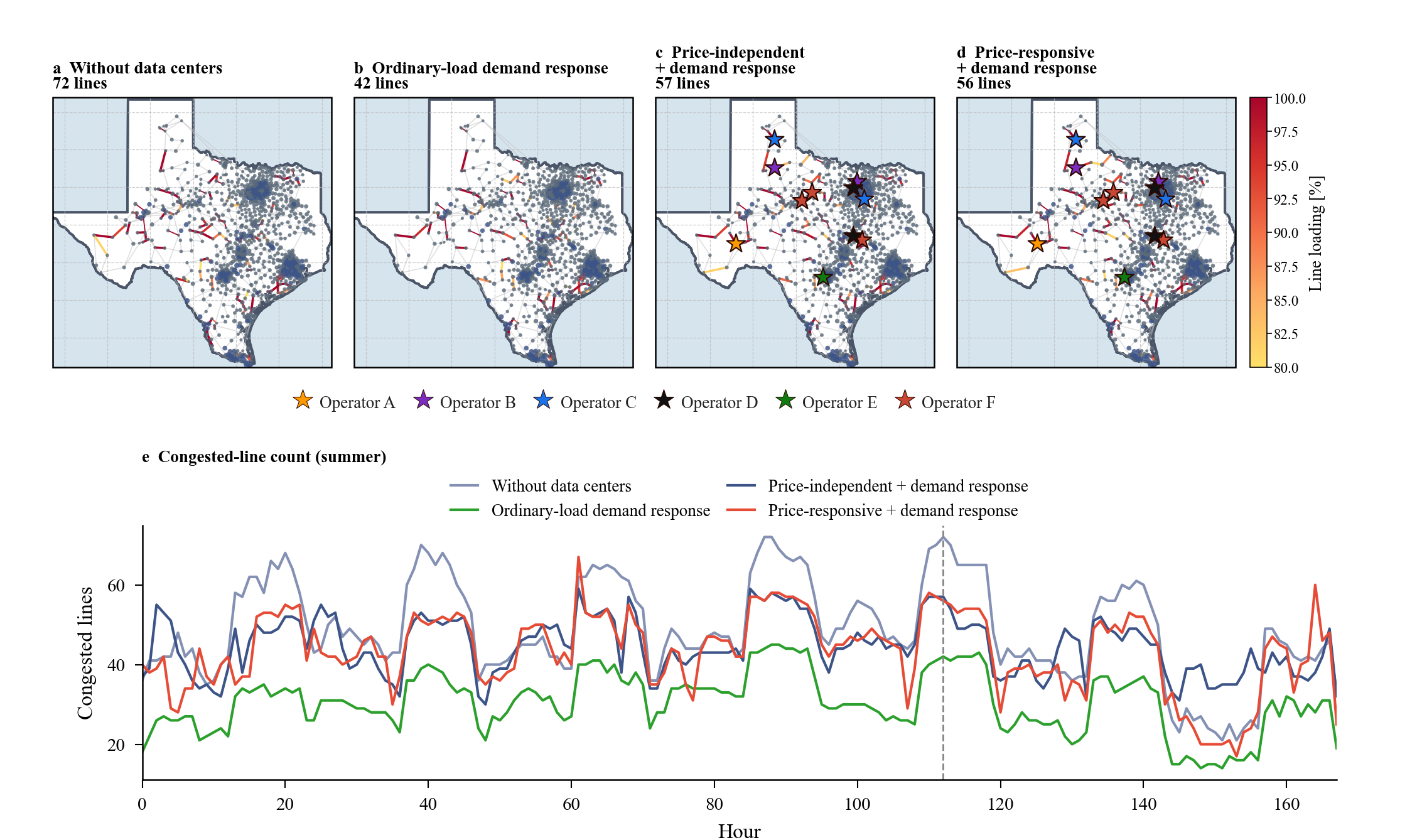}
\caption{\textbf{Transmission loading with ordinary-load demand response.}
\textbf{a--d}, Summer transmission line loading without explicitly modeled data centers or demand response, with ordinary-load demand response alone, and with residual ordinary-load demand response alongside price-independent or price-responsive data center demand, respectively.
Ordinary-load demand response is represented by cyclic storage units and jointly optimized with network dispatch. In \textbf{c,d}, part of the ordinary-load response capacity is displaced to accommodate data center capacity.
\textbf{e}, Corresponding hourly counts over the 168-hour window.}
\label{fig:supp-ordinary-dr}
\end{figure*}

\begin{figure}[!htbp]
\centering
\includegraphics[width=\textwidth]{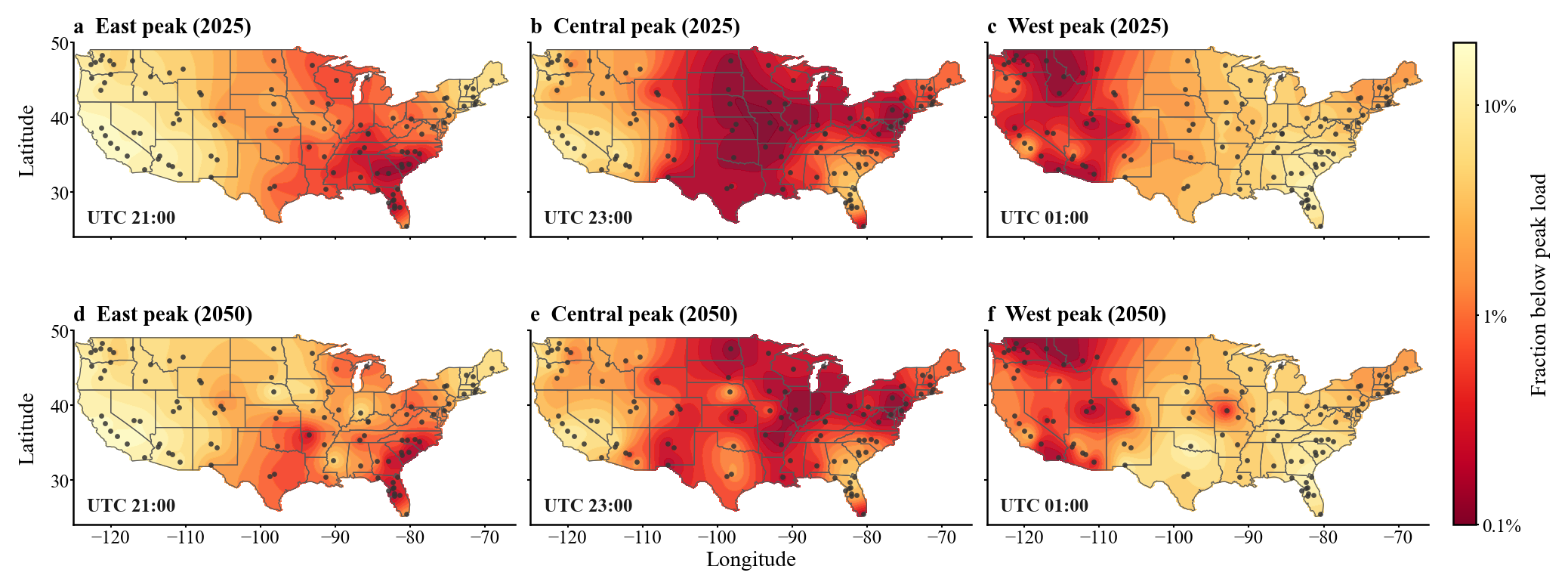}
\caption{\textbf{Illustrative regional electricity demand under a 2050 data center growth scenario.}
\textbf{a--c}, Background demand in 2025 at 21:00, 23:00 and 01:00 UTC as shown in Fig.~1.
\textbf{d--f}, Corresponding demand after adding computing demand scaled to the 2050 projection from the Annual Energy Outlook 2026~[1].}
\label{fig:regional_demand_fullflex}
\end{figure}

\begin{figure*}[h]
\centering
\includegraphics[width=\textwidth]{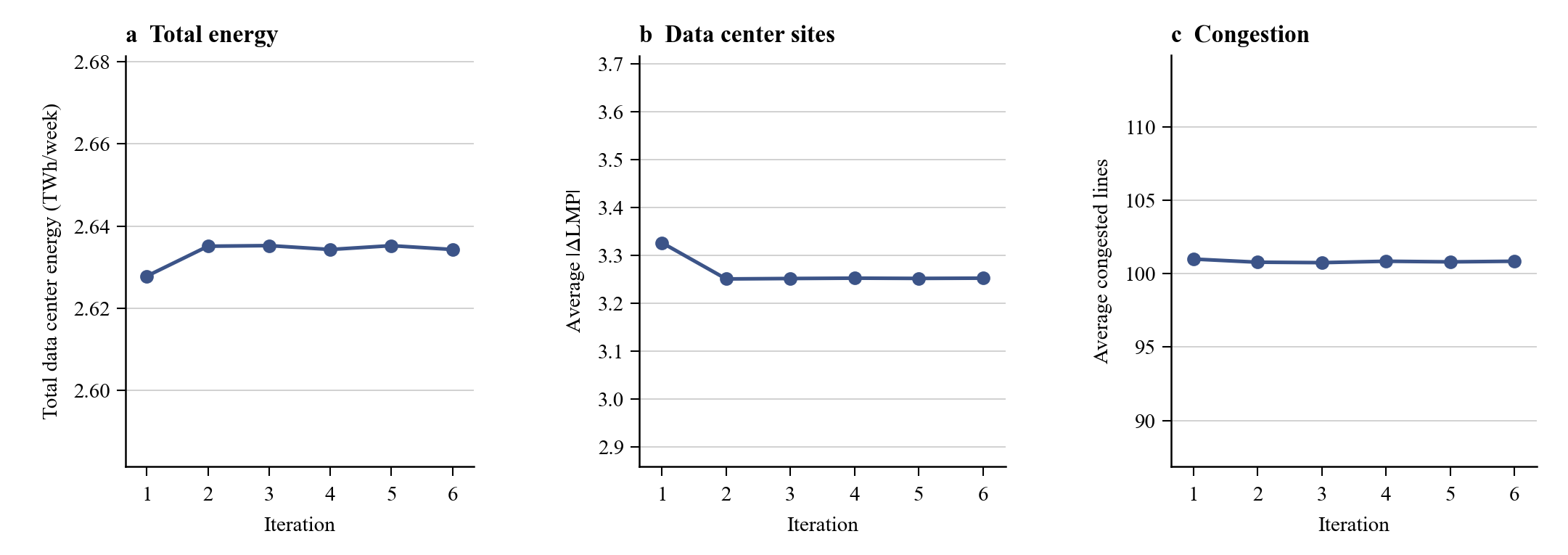}
\caption{\textbf{Summary metrics across six sequential price updates in the continental price-feedback analysis.}
\textbf{a}, Total modeled data center energy consumption.
\textbf{b}, Mean absolute nodal price difference between modeled data center sites.
\textbf{c}, Mean number of congested lines. The summary metrics change little after the second update.}
\label{fig:supp-price-iteration}
\end{figure*}

\begin{figure*}[h]
\centering
\includegraphics[width=\textwidth]{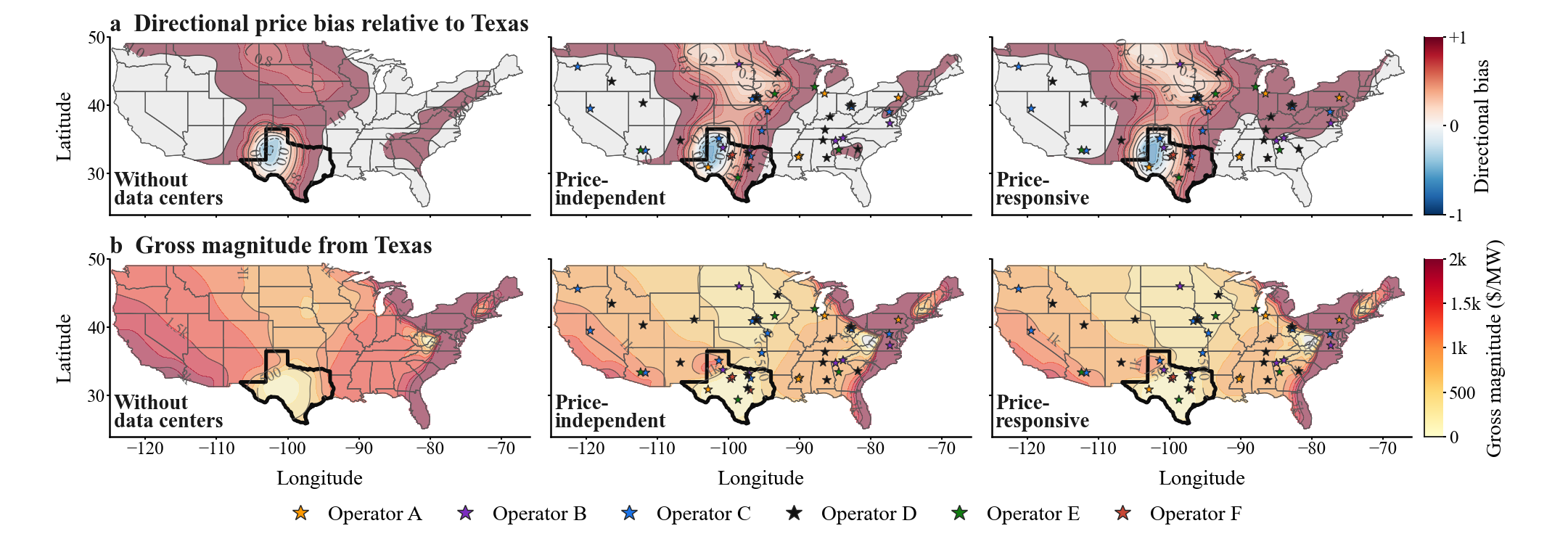}
\caption{\textbf{Winter spatial differences in dispatch-derived marginal prices.}
Without data center, price-independent, and price-responsive cases corresponding to Fig.~6.}
\label{fig:supp-usa-winter}
\end{figure*}

\begin{figure*}[h]
\centering
\includegraphics[width=\textwidth]{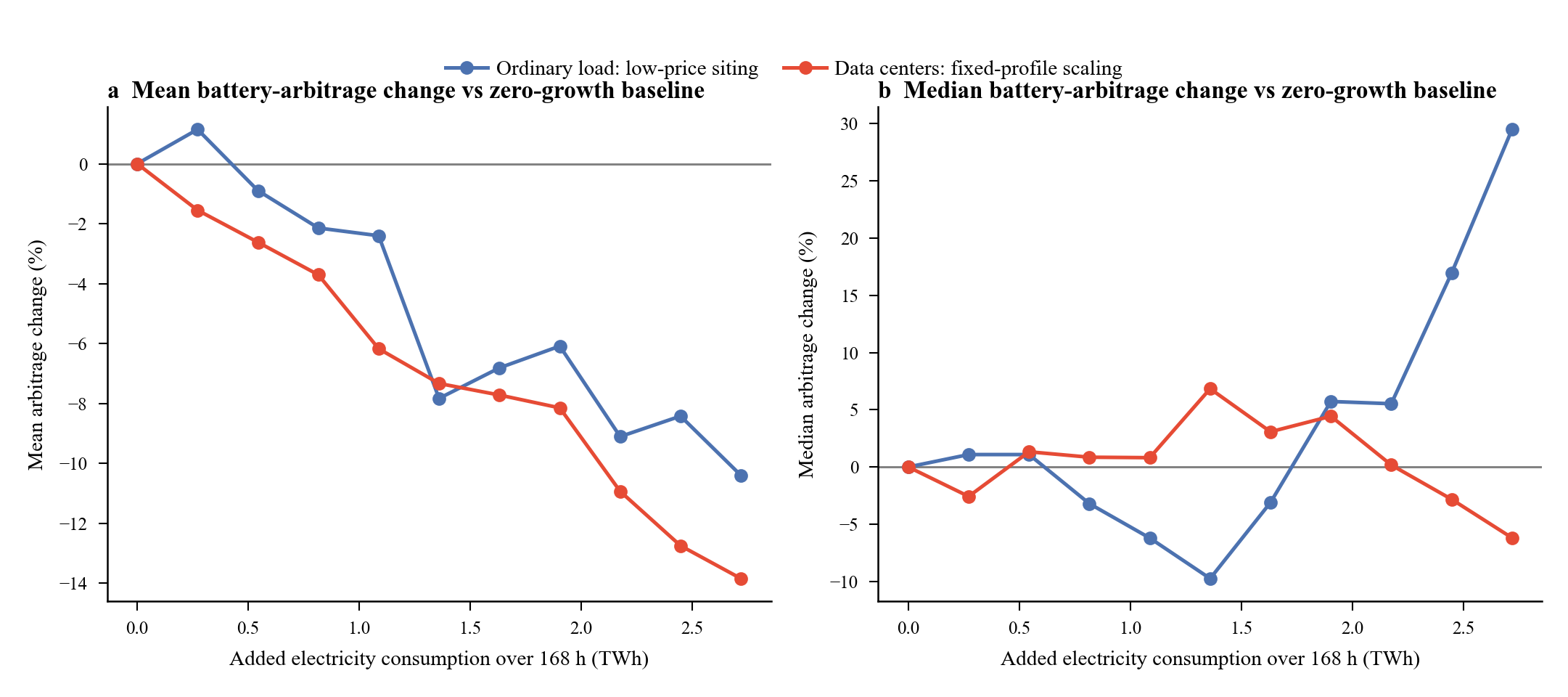}
\caption{\textbf{Battery arbitrage opportunities under ordinary-load and data center growth.}
\textbf{a,b}, Changes in mean and median battery energy-arbitrage value relative to the baseline without added load.
Ordinary-load growth follows a fixed demand shape and is progressively assigned to low-price nodes; data center growth scales existing price-responsive site demand profiles without re-optimizing scheduling.
Both scenarios add the same electricity consumption at each stage and are evaluated over the same 168-hour summer window.}
\label{fig:supp-load-growth}
\end{figure*}

\begin{figure*}[h]
\centering
\includegraphics[width=\textwidth]{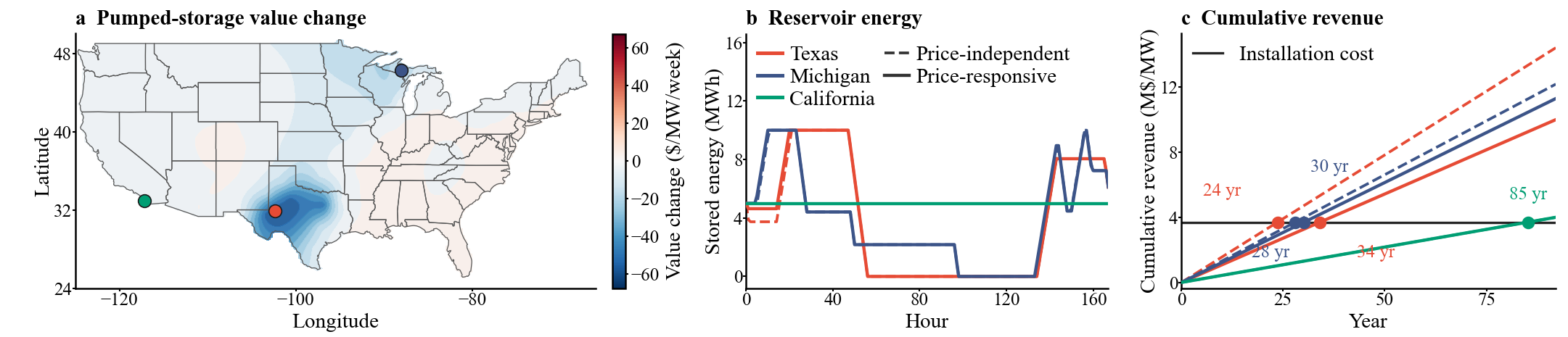}
\caption{\textbf{Pumped-storage hydropower arbitrage under alternative computing schedules.}
\textbf{a}, Change in arbitrage value, calculated as price-responsive minus price-independent.
\textbf{b}, Stored energy profiles.
\textbf{c}, Illustrative cumulative revenue relative to installation cost.
Dashed and solid lines indicate price-independent and price-responsive computing demand, respectively.}
\label{fig:supp_hydro_arbitrage}
\end{figure*}

\clearpage
\section*{Supplementary References}

\noindent
[1] U.S. Energy Information Administration.
\textit{Annual Energy Outlook 2026}
(U.S. Energy Information Administration, Washington, DC, 2026).
\url{https://www.eia.gov/outlooks/aeo/}.

\end{document}